\documentclass[a4paper]{article}

\usepackage{graphicx}
\usepackage{dcolumn}
\usepackage{bm}
\usepackage{siunitx}

\usepackage[utf8]{inputenc}
\usepackage[T1]{fontenc}
\usepackage{mathptmx}
\usepackage{etoolbox}
\usepackage{xcolor}
\usepackage{makecell}
\usepackage{float}
\usepackage{amsmath}
\usepackage{array, booktabs, tabularx}
\usepackage{a4wide}

\title{Combined dynamic-kinematic validation of droplet-wall impact modeling}
\author{D. Zharikov\footnote{Moscow Aviation Institute (National Research University), Moscow 125993. Email: dimazharikov10@gmail.com}, M. Piskunov\footnote{School of Energy and Power Engineering, Tomsk Polytechnic University, Tomsk 634050} and D. Kolomenskiy\footnote{Center for Materials Technologies, Skolkovo Institute of Science and Technology, Moscow 121205}}

\date{\today}

\begin{document}
	
\maketitle

\begin{abstract}
Many numerical studies validate droplet wall impact using only maximum spreading diameter, yet this metric alone cannot ensure correct droplet dynamics.
We present a combined dynamic contact angle (DCA) model that merges the geometric accuracy of the generalized Hoffman-Voinov-Tanner law with the kinematic consistency of a Hoffman function-based approach, improving predictions of droplet spreading and receding.
We simulate water-glycerol droplet impact on sapphire glass at Weber numbers 20-250 and assess both contact angle formulations.
Simulated radial velocity fields are processed in Python using SciPy and compared with Particle Image Velocimetry measurements in the longitudinal section of the spreading droplet.
The Hoffman function-based model captures the main droplet kinematic trends and provides more consistent receding dynamics.
The generalized Hoffman-Voinov-Tanner law matches the maximum spreading diameter within 7\%.
However, during receding, it shows a median absolute error in radial velocity up to three times higher than that of the Hoffman function-based solution.
Average radial velocity and spreading velocity can differ from experimental trends even when maximum spreading is reproduced.
These findings support validation combining geometric and kinematic metrics and motivate the combined model for predicting spreading and receding.
Using the maximum spreading factor $\beta_{max}$ as the ratio of the maximum spreading diameter over the initial droplet diameter and the characteristic capillary number $Ca_{char}$ defined from the mean internal horizontal velocity at 300 \si{\micro\meter} above the substrate, we introduce a $(\beta_{max},\,Ca_{char})$ diagram to relate spreading characteristics to internal flow dynamics.
We hypothesize that, given sufficient data, the contact-line geometry may be used to estimate internal kinematics.
\end{abstract}

\section{Introduction}

The process of droplet impact with a solid substrate \cite{shah2024drop} plays an important role in a wide range of applications, such as high-temperature surface cooling \cite{liang2017review}, spray coating \cite{wang2013drop}, 3D-bioprinting \cite{ramadan20213d}, inkjet printing \cite{modak2020drop}, precision agricultural spraying and crop protection \cite{wang_agri}, anti-icing surface engineering \cite{jiang2024energy} and drop-on-demand metal additive manufacturing \cite{GILANI2021102402}.
The dynamics of droplet impact on solid substrates are governed by a complex interplay of inertial, viscous, capillary, and thermal forces, each contributing to different phases of droplet dynamics \cite{josserand2016drop, yarin2006drop}. 
All possible outcomes of droplet collision are highly sensitive to parameters such as droplet velocity, surface roughness, wettability, and ambient conditions \cite{josserand2016drop}.

Computational fluid dynamics (CFD) simulation of these processes remains a challenging and actively investigated topic, since it must capture multiphase flow and moving interfaces. 
Reliable predictions require careful numerical setup and validation against experiments. 
In particular, validation should include not only droplet shape and spreading, but also internal flow metrics.

A great number of CFD simulations \cite{malgarinos2014vof, zhang2016spreading, bordbar2018maximum, dwiyantoro2013capillary, zheng2023computational} of droplet impact on a solid surface involve validating through maximum droplet spreading diameter.
However, it is questionable whether focusing solely on geometric parameters provides an accurate representation of droplet impact dynamics.
Furthermore, important kinematic parameters of the liquid, such as the internal velocity fields, radial velocity and droplet spreading velocity are not considered in the vast majority of the earlier simulations.

More recently, Shu et al. \cite{SHU2026105449} investigated the impact dynamics of droplets with initial angular velocity on superhydrophobic surfaces. 
They used numerical simulations to explore the effect of droplet rotation on impact behavior.
They found that increasing the initial angular velocities leads to greater centrifugal forces, which enhances droplet spreading and reduces contact time. 
To validate the numerical model, they used geometrical and dynamic parameters including the spreading factor, droplet height, droplet shape evolution over time, rotation angle, and contact time.

Ye et al. \cite{fluids10050131} used the CFD simulation to find out that a dynamic contact angle model accurately predicts the paint droplet impact on dry surfaces.
At the same time, the impact on wet surfaces results in the creation of the craters, whose size strongly correlates with the Reynolds number.
The researchers developed a model to simulate the motion of pigment flakes within the droplet, which is crucial for understanding and improving the final quality of metallic paint finishes.
The numerical results were validated against experimental data by comparing the droplet contour evolution, droplet spread factor and height ratio.

Wang et al. \cite{wang2025study} studied the spreading of liquid droplets on coal dust surfaces to optimize spray dust removal.
Increasing coal-dust surface roughness promotes spreading, yielding a larger maximum spread ratio.
The authors employed high-speed imaging to capture the transient wetting behavior of droplets impacting coal dust and built a two-phase CFD model simulating droplet-particle interactions, including spreading, bouncing, and fragmentation.
The experimentally observed dynamics were then used to validate the CFD model, which demonstrated high accuracy in reproducing the droplet impact and spreading behavior.

The combination of CFD simulations and Particle Image Velocimetry  \cite{meinhart1999piv} (PIV) is a powerful yet challenging approach to studying fluid mechanics.
PIV provides highly detailed velocity field measurement, which can help validate numerical models.
On the other hand, coupling these techniques is not trivial.
The two approaches are affected by fundamentally different sources of uncertainty.
PIV is limited by measurement error and experimental biases, whereas CFD accuracy depends on modelling setup and discretization techniques.

Gultekin et al. \cite{gultekin2020piv} investigated the internal flow dynamics of single and two-component droplets impacting on the sapphire glass solid surface using PIV.
The velocity fields revealed a nonlinear behavior at the lamella edges for experiments with moderate Weber numbers, while test with high Weber numbers showed more linear profiles.
Subsequently, they calibrated CFD simulations using the experimental spreading diameters and PIV velocity measurements to match them.
They compared these characteristics with those obtained by an analytical model.
The linear model agreed well with the linear parts of the velocity profiles in the early stage of spreading, while CFD simulations reproduced both radial velocity distributions and droplet shapes.
The authors also matched the radial velocity distributions, the non-dimensional spreading diameter over time, and the shape of the droplet.

Erkan  \cite{erkan2019full} described the velocity fields inside water droplets impacting heated sapphire surfaces using time-resolved PIV and shadowgraph imaging.
Validation of velocity measurements included repeated trials, comparison of PIV data with spreading velocities from shadowgraph images. 
For the unheated surface impacts, the experimental results were compared with data produced by analytical models of viscous spreading film, which captured spreading velocities but failed to reproduce the nonlinearity of the velocity profiles.
CFD simulations reproduced droplet deformation and spreading velocities, with acceptable deviations at early stages and at high Weber numbers.

Schubert et al. \cite{schubert2024micro} explored the dynamics of droplet impact on thin liquid films with varying viscosities by combining micro-PIV measurements, analytical modeling, and direct numerical simulations (DNS).
The velocity fields inside the spreading lamella were resolved with high accuracy.
This results in a linear radial increase consistent with an axisymmetric stagnation point flow and an exponential temporal decay linked to the pressure force exerted by the droplet.
Validation of the CFD simulations was performed through a direct comparison of micro-PIV velocity data with DNS predictions, analytical models, and literature results.
The study highlights the transition between inertial and viscous regimes and establishes a robust framework for validating velocity field measurements in droplet impact studies.

Paturalski and Tomaszewski \cite{paturalski2025application} employed PIV as a benchmark for CFD simulations of a pressure-swirl atomizing nozzle.
High resolution measurements of peak velocities, coherent vortex structures and pronounced turbulence intensity were validated against CFD simulations using the $k-\omega$ turbulence model coupled with the Volume of Fluid method.
Although peak velocities and swirl patterns were accurately simulated, there were 10-15\% deviations in downstream decay due to model assumptions and PIV limitations.
The integration of 3D-scanned nozzle geometry and statistical validation metrics confirmed that PIV is a precise method for CFD benchmarking.

The goal of this study is to develop and to test a combined dynamic-kinematic validation method to improve the fidelity of droplet-impact modeling, while also demonstrating that relying solely on the maximum spreading diameter is an insufficient validation criterion.
We assume that rigorous validation requires consideration of both the deviation of the maximum droplet spreading diameter and the flow kinematic characteristics. 
These characteristics are quantified using internal-velocity metrics, including PIV-based velocity fields, cross-section-averaged radial velocity, and contact-line spreading velocity. 
This integrated design enables a direct test of whether validation based solely on the maximum spreading diameter is sufficient to capture the hydrodynamics of the impact process.
In contrast to other studies, which primarily validated a CFD model using only geometric parameters, our numerical study evaluates both geometric and kinematic characteristics of droplet impact dynamics.

We believe that the proposed method is more robust compared to the geometry-only validation methods.
The method is verified using the water-glycerol droplet on sapphire glass at impact Weber numbers ($We_{imp}$) in a range of 20-250 and droplet impact velocities $U_{0}$ in a range of 0.63-2.08 m/s.
Two dynamic contact angle models are evaluated: the generalized Hoffman-Voinov-Tanner law and the Hoffman function.
To unite the advantages of both models, a combined DCA (Dynamic Contact Angle) model is proposed.

\section{Materials and Methods}

\subsection{Experimental Methods}

We use data from earlier experiments by Ashikhmin et al. \cite{ASHIKH_COAL} on water-glycerol droplet collisions against a sapphire glass substrate. 
The droplets contained 60 wt\% glycerol, 40 wt\% water, and titanium dioxide particles with the concentration of 1 $g/l$. 
The main parameters of the setup are listed in Table \ref{tab:liquid_properties}.

\begin{table*}[t]
\centering
\caption{Conditions of the experiment} 
\label{tab:liquid_properties}
\renewcommand{\arraystretch}{1.5} 
\begin{tabular*}{\textwidth}{@{\extracolsep{\fill}}ccccc} 
\hline \hline
Temperature (K) &
Density, $\rho$ (kg/m$^3$) &
Dynamic viscosity, $\mu$ (Pa~s) &
Surface tension, $\sigma$ (N/m) &
Static contact angle, $\theta_0$ \\
\hline
293.15 & 1154 & $10.8 x 10^{-3}$ & 0.06058 & 46.5$^\circ$ \\
\hline \hline
\end{tabular*}
\end{table*}

The impacts were filmed using high-speed videography, then the internal convective flow velocities were evaluated by particle image velocimetry (PIV) within a horizontal plane at a constant height ($250 \pm 50$ \si{\micro\meter}) above the substrate.
The pixel size of PIV was 0.114 mm.
The laser beam had a thickness of 200 \si{\micro\meter}.

The impact dynamics were examined across a range of $We_{\text{imp}} \in \{20, 80, 250\}$, which characterize the ratio of inertial to surface tension forces.
The impact Weber number is defined as $We_{imp} = \rho U_{0}^2 D_0/ \sigma$, where $\rho$ is density, $U_{0}$ is an impact velocity, $D_0$ is initial droplet diameter, and $\sigma$ is surface tension.

The velocity of the droplet at the moment of impact was 0.63 m/s, 1.17 m/s, and 2.08 m/s, and the systematic measurement error was 0.1 m/s. 
Each experiment with corresponding impact velocity was repeated three times.
The diameter of the droplet remained constant and equal to 2.9 ± 0.05 mm throughout the entire time interval from detachment to collision.
The correspondence between $We_{imp}$ and the velocities of droplet impact against the substrate $U_0$ is presented in the Table \ref{tab:we_u_corresp}.

\begin{table}[ht] 
\centering
\caption{The correspondence between $We_{imp}$ and $U_0$ in the experiments.}
\label{tab:we_u_corresp}
\renewcommand{\arraystretch}{1.5}
\begin{tabular*}{\columnwidth}{@{\extracolsep{\fill}}cccc@{}}
\hline \hline
$We_{imp}$ & 20 & 80 & 250 \\
\hline
$U_0$, m/s & 0.63 & 1.17 & 2.08 \\
\hline \hline
\end{tabular*}
\end{table}

\subsection{Governing equations}

A laminar flow model was used to describe the fluid flow. 
The motion of the fluid is governed by the incompressible Navier-Stokes equations. 
A single set of equations is solved for the entire domain.
The governing equations are presented below.

The continuity equation takes the form
\begin{equation}
\nabla \cdot \mathbf{u} = 0.
\end{equation}

The momentum equation is
\begin{eqnarray}
\frac{\partial (\rho \mathbf{u})}{\partial t}
    + \nabla \cdot (\rho \mathbf{u} \mathbf{u}) \\
    =-\nabla p + \nabla \cdot \left[ \mu \left( \nabla \mathbf{u} + \nabla \mathbf{u}^T \right) \right] + \rho \mathbf{g} + \mathbf{F}_{\sigma},
\end{eqnarray}

 where $\mathbf{u}$ is the velocity vector, $\rho$ is the density, $p$ is the pressure, $\mu$ is the dynamic viscosity, $\mathbf{g}$ is the gravitational acceleration vector, $\mathbf{F}_{\sigma}$ is the surface tension force.

The Volume of Fluid \cite{hirt1981volume} (VOF) model based on an Eulerian approach was used to simulate the multiphase flow, in which two phases (air and liquid) were considered.

The interface is characterized by the volume fraction of the liquid phase, denoted by $\alpha$, which is defined as
\begin{equation}
    \alpha=
    \begin{cases}
        1 & \textit{in the liquid phase,} \\
        0 < \alpha < 1 & \textit{at the interface,} \\
        0 & \textit{in the gas phase.}
    \end{cases}
\end{equation}

The mixture density (Eq. \ref{eq:mixture_density}) and dynamic viscosity (Eq. \ref{eq:mixture_viscosity}) are computed as
\begin{align}
    \rho & = \alpha \rho_l + (1 - \alpha) \rho_g, \label{eq:mixture_density} \\
    \mu & = \alpha \mu_l + (1 - \alpha) \mu_g. \label{eq:mixture_viscosity}
\end{align}

where subscripts $l$ and $g$ denote liquid and gas phases, respectively.

The evolution of the interface is governed by the advection equation
\begin{eqnarray}
     \frac{\partial \alpha}{\partial t}
    + \nabla \cdot (\alpha \mathbf{u}) = 0.
\end{eqnarray}

\subsection{Numerical Methods}

Ansys Fluent 2023 R1 software package was used to simulate the collision of a droplet with a substrate.
A 2D axisymmetric formulation was used, taking advantage of the rotational symmetry of a spherical droplet impacting normally against a smooth substrate and thereby reducing the problem to a two-dimensional domain. 
Moreover, no splashing was observed in the experiments, justifying the axisymmetric assumption reducing computational cost relative to 3D simulations.
SIMPLE \cite{patankar2018numerical} (Semi-Implicit Method for Pressure Linked Equations) algorithm was used to couple the pressure and velocity fields when solving the Navier-Stokes equations. 
The pressure gradient is calculated using the least squares method, and the pressure itself is sampled using the PRESTO! (Pressure Staggering Option) algorithm. 
The momentum equations are solved by a second-order upwind scheme. 
The Geo-Reconstruct scheme \cite{youngs1982time} is used for the volume fraction.
To more accurately simulate real-world conditions, gravitational acceleration is applied in the negative y-direction. 
The magnitude is $g = 9.81\ \text{m}/s^2$.

The initial conditions in the simulations (density, dynamic viscosity, surface tension, and static contact angle) are identical to those provided in Ref. \cite{ASHIKH_COAL} for the water-glycerol sample.
To reduce computational cost, the droplet motion is initiated from a height $h_0$ of 3 mm.
The initial droplet velocity $U_{init}$ is calculated as

\begin{equation}
    U_{init}=\sqrt{U_{0}^2-2gh_0}.
    \label{ref:init_vel}
\end{equation}

\begin{figure}[!h]
\centering
\includegraphics[width=0.45\textwidth]{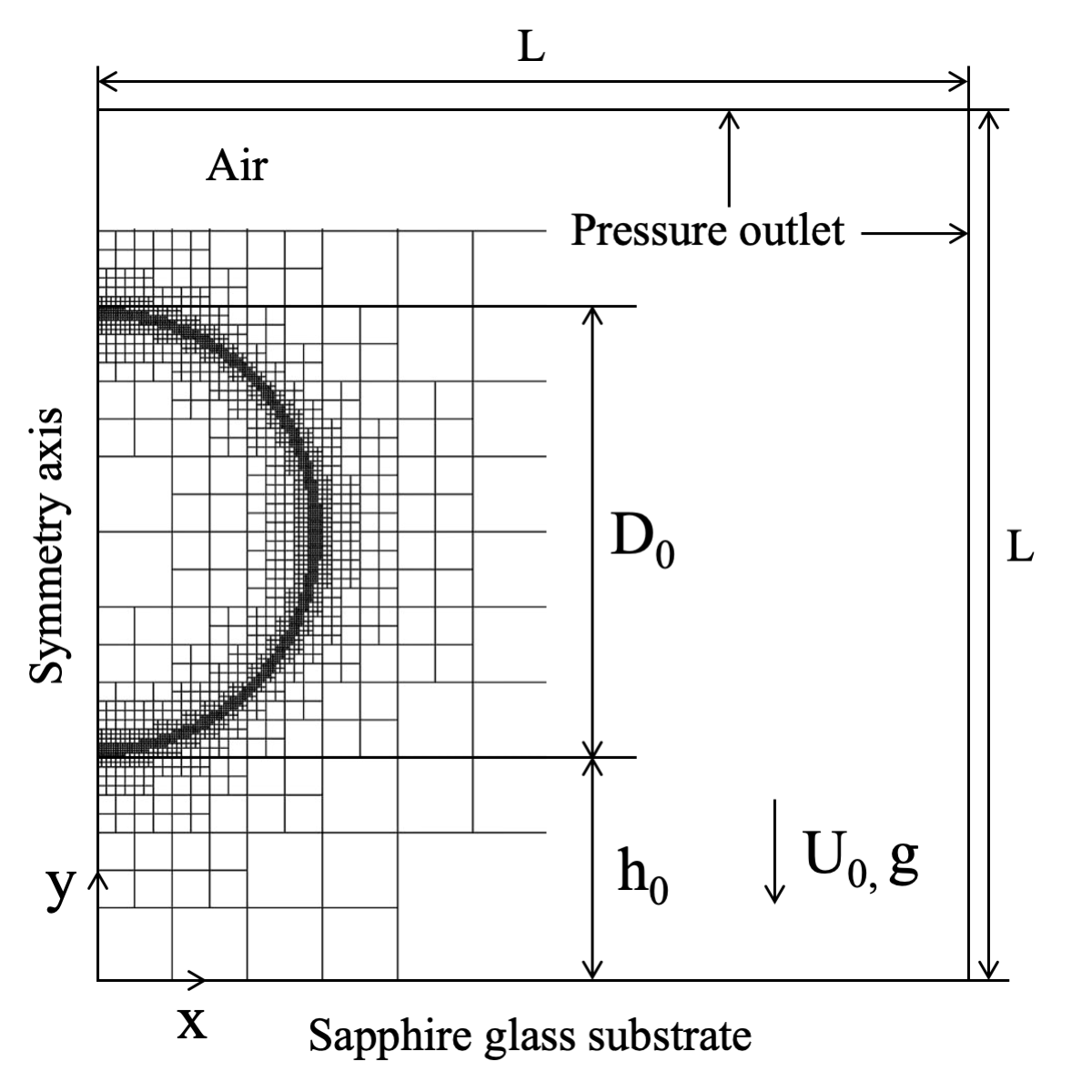}
\caption{\label{fig:comp_mesh}Schematic drawing of the computational domain, geometrical parameters and mesh.}
\end{figure}

The computational domain is a two-dimensional square area with a side length of 40 mm.
The top and right boundaries of the domain are defined as pressure outlets.
The left boundary of the domain at $y=0$ represents the axis of symmetry.
The bottom boundary represents an impermeable wall with no-slip condition.
The above mentioned conditions together with the geometry of the problem and a computational mesh are illustrated in Figure \ref{fig:comp_mesh}.

\subsection{Dynamic Contact Angle Models}

Dynamic contact angle models are crucial for capturing wetting behavior under transient conditions, which strongly affect flow, adhesion, and spreading phenomena.
The models relate the apparent contact angle to the contact-line velocity. 
This relation enables predictive modeling of transient wetting on solids \cite{cox1986dynamics, SEDEV2015661}.
With dynamic contact angle models, simulations can reproduce contact-line hysteresis \cite{snoeijer2013moving}.
They also capture transverse rim instabilities that generate fingering during droplet impact and the associated inhomogeneous internal velocity fields within the drop \cite{VOZHAKOV2025108372, kumar2017internal}.
In this study, two dynamic contact angle models were applied via the user-defined functions.

The first is the generalized Hoffman-Voinov-Tanner (HVT) law \cite{hoffman1975study} for moving contact line dynamics that is defined by equation  

\begin{equation}
    \theta_D^3 - \theta_0^3 = \kappa Ca,
    \label{eq:generalized}
\end{equation}

where $\theta_0$ is the static contact angle and $\theta_D$ is the dynamic contact angle, $\kappa$ is an empirical parameter of the model. 
The capillary number is defined as $Ca = \mu U_{spr} / \sigma$, where $U_{spr}$ represents the contact line velocity.
The model is valid only in cases where the static contact angle is finite and greater than zero.

In the second model, the contact angle is calculated using the Hoffman function \cite{kistler1993hydrodynamics}.
For the advancing mode, the contact angle is calculated as

\begin{equation}
    \theta = f_H\left(Ca + f_H^{-1}(\theta_0)\right), \quad Ca > 0.
    \label{ref:hoff_adv}
\end{equation}

For the receding mode it is
\begin{equation}
    \theta = \frac{\theta_0}{\pi - \theta_0} \left[\theta_0 - f_H\left(-Ca + f_H^{-1}(\theta_0)\right)\right] + \theta_0, \quad Ca < 0.
    \label{ref:hoff_rec}
\end{equation}

The Hoffman function $f_H(x)$ is defined as

\begin{equation}
f_H(x) = \arccos\Bigl(1 - 2 \tanh\Bigl(5.16 \Bigl(\frac{x}{1 + 1.31 x^{0.99}}\Bigr)^{0.706}\Bigr)\Bigr)
\end{equation}

\subsection{Adaptive Mesh Refinement and Mesh Convergence}

The adaptive mesh refinement \cite{van2018towards} (AMR) criterion was applied to refine the computational mesh based on the gradient of the VOF field, specifically targeting the water phase to accurately capture the interface between the droplet and the air phase. 
This criterion refines cells where the scaled gradient of the VOF field exceeds a threshold value of 0.06, with the gradient itself scaled by the global maximum to ensure refinement. 

To prove mesh independence, multiple simulations of droplet-wall impact at $We_{imp}=20$ using the Hoffman function-based dynamic contact angle were performed with different adaptation levels.
The average velocity ($v_{avg}$) in the longitudinal direction at a height of $h_{l}=300$ \si{\micro\meter} was chosen as the mesh convergence criterion.
Velocity values were processed using the procedure described in Section \ref{subsec:velocity-preproc}.
Different numbers of adaptation levels with various mesh parameters were tested.
The characteristics of the trials with different adaptation levels are presented in Table \ref{tab:mesh_stats}.

\begin{table}[H]
\centering
\renewcommand{\arraystretch}{1.5}
\caption{Mesh adaptation statistics}
\label{tab:mesh_stats}
\begin{tabular*}{\columnwidth}{@{\extracolsep{\fill}}ccccc@{}}
\hline \hline
\makecell[l]{Number of \\ adaptation levels} &
Cells &
Faces &
Nodes &
\makecell[l]{Minimum \\ face size, m} \\ \hline
3 & 1828 & 3784 & 1957 & 0.000183 \\
4 & 2074 & 4325 & 2252 & 0.000092 \\
5 & 2644 & 5557 & 2914 & 0.000046 \\
6 & 3700 & 7855 & 4156 & 0.000023 \\ \hline \hline
\end{tabular*}
\end{table}

\begin{figure}[t]
\centering
\includegraphics[width=0.45\textwidth]{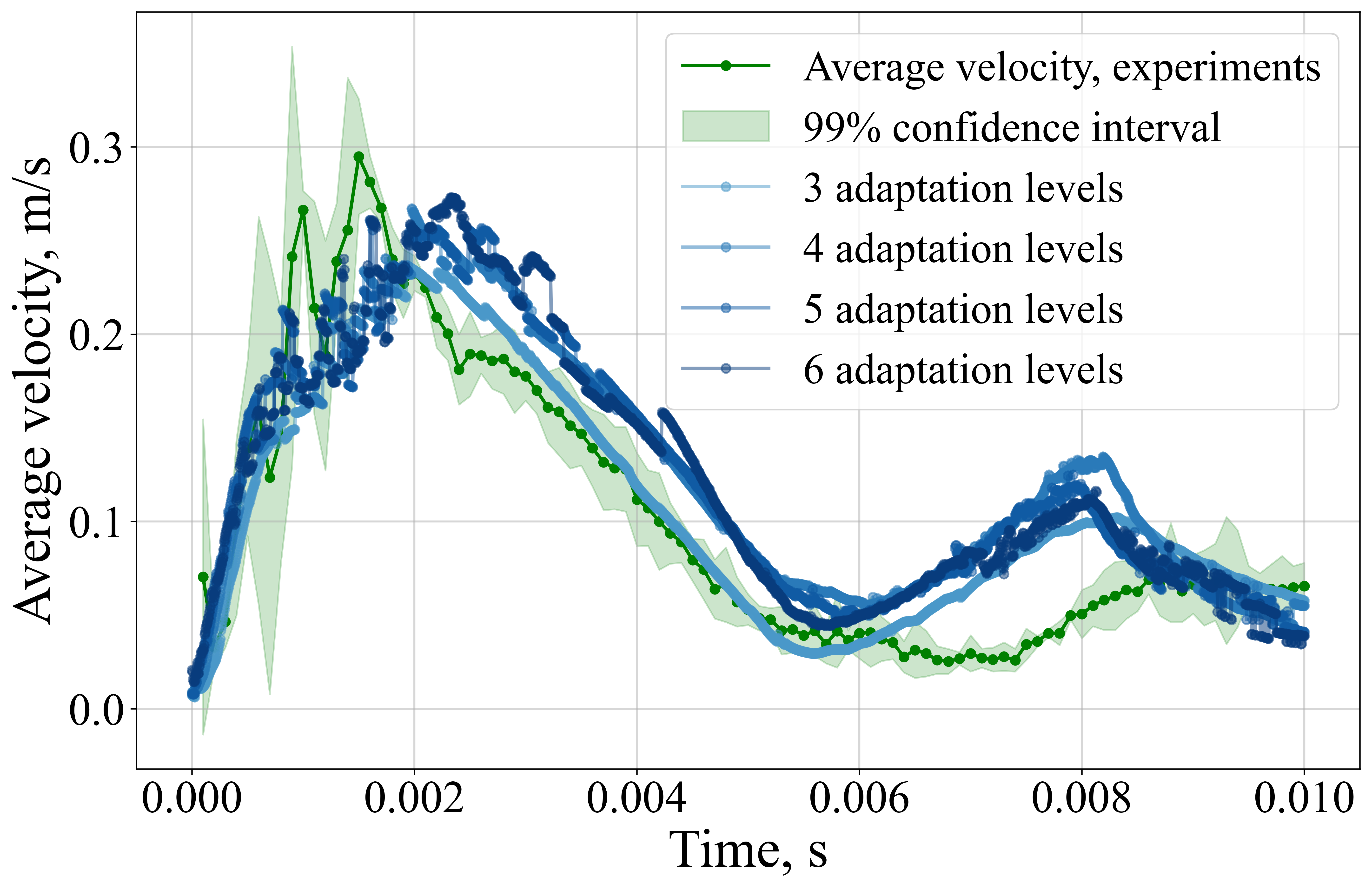}
\caption{\label{fig:conv_mesh} Average values of internal flow velocities for different adaptation levels.}
\end{figure}

The results of the calculations for different adaptation levels are presented in Fig. \ref{fig:conv_mesh}.
Minimal discrepancies in average velocity are observed across different meshes for varying adaptation levels. 
At an adaptation level of 5, an optimal balance between solution accuracy and computational complexity is achieved.
Thus, this level is selected for implementation.
The chosen adaptation level ensures mesh-independent results. 

\subsection{Data interpolation}

For a consistent comparison between experimental and numerical velocity data, the original radial velocity values are preprocessed with multiple interpolation methods. 
These interpolation schemes yield smooth and numerically robust reconstructions of the discrete velocity measurements.

Linear and cubic spline interpolation methods were applied to the velocity values.
The type of interpolation and the exact data processing are described in the next section.
All interpolations were applied with the SciPy \cite{2020SciPy-NMeth} Python library.

\subsection{Velocity values preprocessing}
\label{subsec:velocity-preproc}

When comparing velocity fields from CFD simulations and PIV experiments, it should be noted that PIV yields spatially averaged velocities over the interrogation window, whereas CFD provides mesh-dependent numerical values. 
Hence, CFD data should be interpolated to a comparable grid and filtered prior to quantitative comparison.

To calculate average velocity values in the longitudinal section of droplets for every timestep, some transformations were applied for the data obtained from the simulation.
The preprocessing steps for velocity values include the following:

\begin{enumerate}
    \item Linear interpolation of velocity values on a uniform mesh.
    \item Data filtration using the Fast Fourier Transform \cite{brigham1988fast} (FFT).
    \item The replacement of velocity values in cells further than 75\% of the maximum distance from the symmetry axis with decreasing cubic spline interpolation.
    \item Velocity weights calculation.
    \item Average weighted velocity calculation.
\end{enumerate}

To eliminate high-wavenumber noise components of values obtained from a uniform mesh, a wavenumber-domain filtering approach based on the FFT is employed.
In the wavenumber domain, a signal is represented as a superposition of harmonic components, each of which is characterized by its amplitude and wavenumber. 
To suppress high-wavenumber components associated with noise, a Gaussian filter is applied as

\begin{equation}
    f(w)=exp(-a(k/k_{max})^2),
\label{ref:gauss_filter}
\end{equation}

where $k$ is the current wavenumber, $k_{max}$ is the maximum wavenumber determined by the spatial resolution of the data ($k_{max}=\pi / H_{PIV}$), the parameter $a$ controls the intensity of suppression ($a=0.1$).
The filtered signal was processed with the inverse FFT.

To compare the simulation values with the results obtained in the experiments, the velocity values further than 75\% of the distance from the symmetry axis were replaced with cubic spline-interpolated $u_{spline, i}$ values weighted by the ratio $u_{spline, i}/u_{uniform, i}$, where $i$ is the index of a cell on a uniform mesh. 
These values were modified to account for PIV measurement errors, as described in Section \ref{subsec:velocity-original}.

The average velocity values for every timestep $t$ were calculated as mean preprocessed radial velocity at a height corresponding to experimental data. 
It is defined as
\begin{equation}
    v_{avg, t}=\frac{1}{\pi r_N^2} \left( \sum_{i=1}^N 2 \pi w_i r_i \, dr \right),
\label{ref:average_velocity_formula_updated}
\end{equation}

where $r_i$ is the distance from the axis of symmetry, $dr$ is the step of a uniform mesh, $r_N$ is the maximum distance from the axis of symmetry and $w_i$ is modified (weighted) radial velocity value
$$
{
w_i(r_i) =
\begin{cases}
    u_{\textrm{uniform}, i}, & \textrm{if} r_i < \frac{3}{4}r_N, \\
    u_{\textrm{spline}, i} \frac{u_{\textrm{spline}, i}}{u_{\textrm{uniform}, i}}, & \textrm{if } r_i \geq \frac{3}{4}r_N.
\end{cases}
}
$$

\begin{figure*}
\centering
\includegraphics[width=1\textwidth]{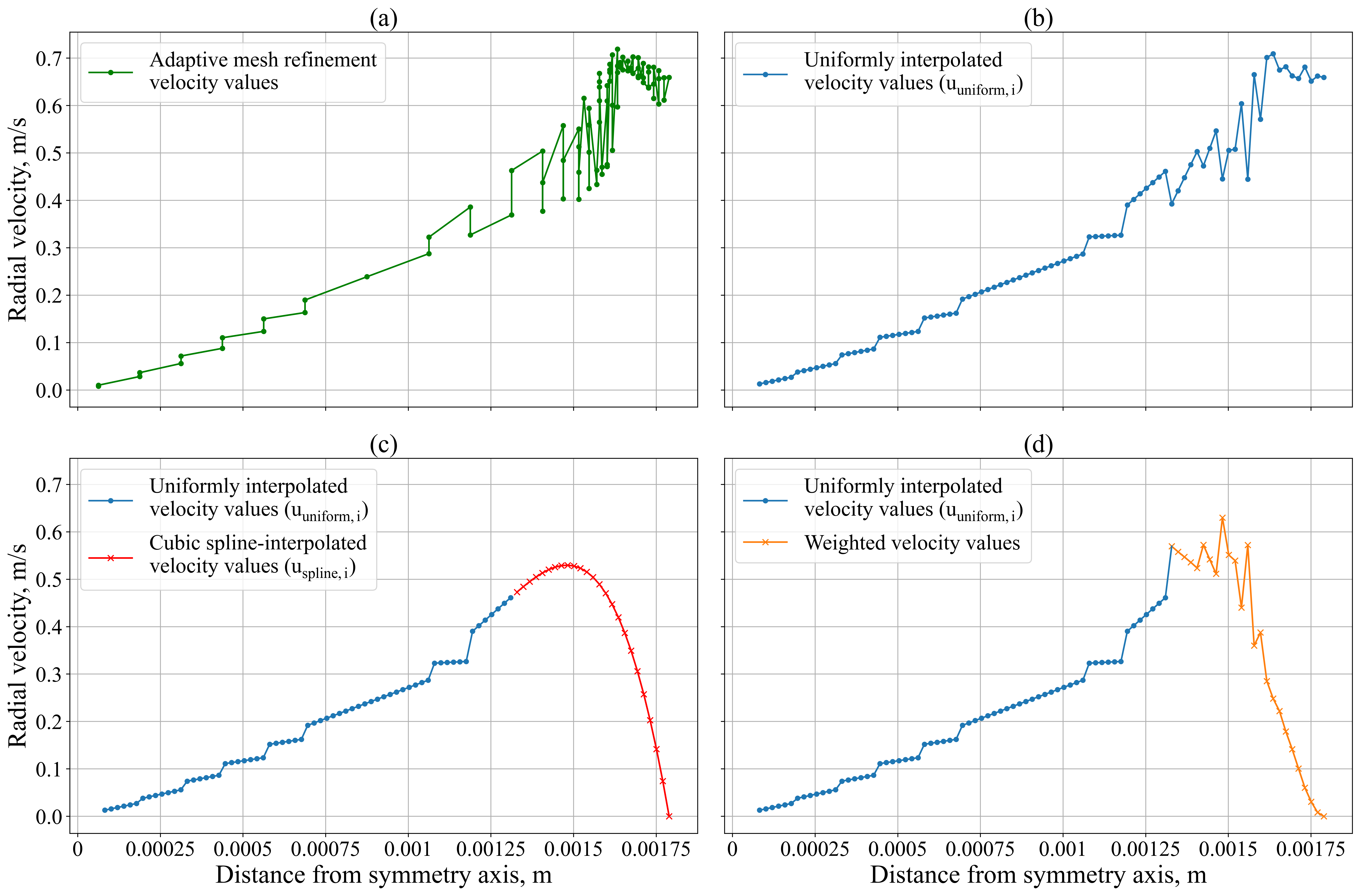}
\caption{\label{fig:spline_plot} Radial velocity data preprocessing: (a) - original adaptive mesh refinement velocity values, (b) - velocity values interpolated on a uniform mesh ($u_{\textrm{uniform}, i}$), (c) - velocity values interpolated on a uniform mesh ($u_{\textrm{uniform}, i}$, blue) with cubic spline interpolated values ($u_{\textrm{spline}, i}$, red) further than 75\% of maximum distance from symmetry axis, (d) - velocity values interpolated on a uniform mesh ($u_{\textrm{uniform}, i}$, blue) with weighted velocity values ($u_{\textrm{spline}, i} \frac{u_{\textrm{spline}, i}}{u_{\textrm{uniform}, i}}$, orange).}
\end{figure*}

The example of data processing for case $We_{imp}=20$ at $t=0.003885$ s is presented in Fig. \ref{fig:spline_plot}, where Fig. \ref{fig:spline_plot} (a) shows the initial data, while Figures \ref{fig:spline_plot} (b-d) show consequent steps of data preprocessing.

\section{\label{sec:res_and_disc}Results and discussion}

\subsection{\label{sec:max_drop_spr_diam}Maximum droplet spreading diameter}

The generalized Hoffman-Voinov-Tanner law \cite{hoffman1975study} contains a constant $\kappa$, the values of which were selected empirically in such a way as to best correspond to the maximum spreading diameter in the experiment ($D_{exp}$). 
The value of the constant that produced the closest results ($\kappa=8.78$) was calculated by linear extrapolation of the results with $\kappa=3$ and  $\kappa=5$

\begin{equation}
    \kappa = 3 + 2\frac{D_{exp}-D_3}{D_5-D_3},
\label{ref:const_interp}
\end{equation}
and taking the average of multiple simulations with different values of $D_{exp}$.
$D_3$ is the simulation diameter obtained with $\kappa=3$, $D_5$ is the simulation diameter obtained with $\kappa=5$.

A comparison of models by maximum droplet spreading diameter is shown in Table \ref{tab:contact_angle_models}.
The model employing a static contact angle exhibits substantial deviation from experimental data, indicating its insufficient physical validity for the given process. 
Conversely, incorporation of a dynamic contact angle into the model yields high accuracy, with discrepancies from the experimental results remaining below 7\% for the most effective dynamic contact angle model. 
This underscores the critical role of accounting for wetting dynamics and time-dependent contact angle variation in accurately simulating droplet spreading behavior.
A widely accepted benchmark for validating droplet-surface impact models is the maximum spreading diameter of a droplet. 
According to this criterion, the model that uses the generalized Hoffman-Voinov-Tanner law performs better.

\begin{table}[H]
\centering
\renewcommand{\arraystretch}{1.5}
\caption{Comparison of the results on the maximum spreading diameter. Percentage difference from experiments.}
\label{tab:contact_angle_models}
\begin{tabular*}{\columnwidth}{@{\extracolsep{\fill}}lccc@{}}
\hline \hline
Contact Angle Model & $We_{imp}=20$ & $We_{imp}=80$ & $We_{imp}=250$ \\
\hline
Constant Contact Angle & 30.33\% & 14.88\% & 10.64\% \\
\makecell[l]{Generalized\\HVT Law \cite{hoffman1975study}, $\kappa=3$} & 16.16\% & 6.21\% & 10.27\% \\
\makecell[l]{Generalized\\HVT Law \cite{hoffman1975study}, $\kappa=5$} & 12.46\% & 3.46\% & 3.20\% \\
\makecell[l]{Generalized\\HVT Law \cite{hoffman1975study}, $\kappa=8.78$} & 6.30\% & 0.27\% & 1.34\% \\
Hoffman Function \cite{kistler1993hydrodynamics} & 11.57\% & 12.52\% & 6.84\% \\
\hline \hline
\end{tabular*}
\end{table}

\subsection{Average radial velocity without PIV correction preprocessing}
\label{subsec:velocity-original}

\begin{figure*}[t!]
    \centering
    \includegraphics[width=1.0\textwidth]{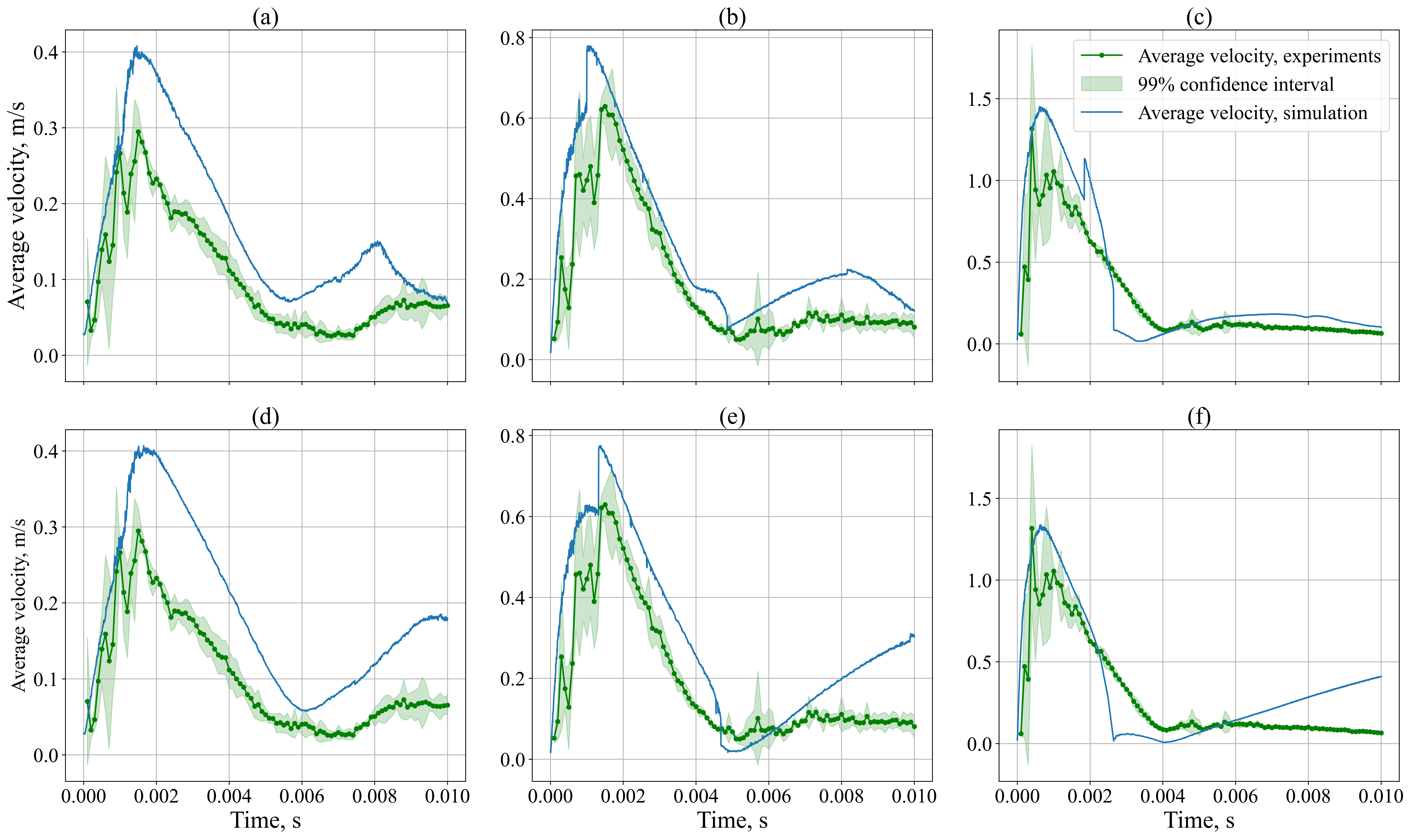}
    \caption{Comparison of the values of the average velocity inside the droplet with experimental data. The numerical results are linearly interpolated without any further modifications. Hoffman function-based simulations are presented at the first row (a, b, c). Simulations with generalized Hoffman-Voinov-Tanner law are presented at the second row (d, e, f). Simulations with $We_{imp}=20$ are presented at the first column of plot (a, d), $We_{imp}=80$ - the second column (b, e), $We_{imp}=250$ - the third column (c, f).}
    \label{fig:velocity_comparison_original}
\end{figure*}

To verify that the kinematic characteristics of the simulations agree with data from the experiments, the radial velocity values were obtained from the centers of the cells at the height of $h_{l}=300$ \si{\micro\meter} above the substrate.
The velocity profiles were linearly interpolated from the non-uniform adaptive mesh to a uniform grid for plotting.
The linearly interpolated numerical radial velocity values are presented in Fig. \ref{fig:velocity_comparison_original}.
This plot visualizes the results for the Hoffman function-based approach (first row of the plot in Fig. \ref{fig:velocity_comparison_original} (a, b, c)) and generalized Hoffman-Voinov-Tanner law (second row in Fig. \ref{fig:velocity_comparison_original} (d, e, f)).
The absolute and relative errors of peak velocity values of the numerical results are presented in Table \ref{tab:compare_orig_data}.

The median absolute error of average velocity values

\begin{equation}
    \text{MedAE} = \text{median}\left( \left| y_1 - \hat{y}_1 \right|, \left| y_2 - \hat{y}_2 \right|, \ldots, \left| y_n - \hat{y}_n \right| \right),
\label{eq:medae}
\end{equation}

between the simulations ($\hat{y}_i$) and the experimental ($y_i$) data for the receding phase (the time period after 8 ms from the droplet-wall contact) is listed in Table \ref{tab:mae_comparison_original}.

\begin{table}[t!]
\centering
\renewcommand{\arraystretch}{1.5}
\caption{A comparison of peak velocity values for different impact velocities. Absolute and percent error of the numerical results.}
\label{tab:compare_orig_data}
\setlength{\tabcolsep}{3pt}
\begin{tabular*}{\columnwidth}{@{\extracolsep{\fill}}l l cc@{}}
\hline \hline
\makecell[l]{$We_{imp}$} &
\makecell[l]{Contact angle\\ model} &
\makecell[l]{Absolute\\ error, m/s} &
\makecell[l]{Relative\\ error, \%} \\
\hline
20 &
\makecell[l]{Hoffman\\ function \cite{kistler1993hydrodynamics}\\ } &
0.11294 & 38.30 \\
20 &
\makecell[l]{Generalized\\ HVT law \cite{hoffman1975study}\\ } &
0.11216 & 38.04 \\
\hline
80 &
\makecell[l]{Hoffman\\ function} &
0.14990 & 23.82 \\
80 &
\makecell[l]{Generalized\\ HVT law} &
0.14568 & 23.15 \\
\hline
250 &
\makecell[l]{Hoffman\\ function} &
0.13428 & 10.19 \\
250 &
\makecell[l]{Generalized\\ HVT law} &
0.02089 & 1.59 \\
\hline \hline
\end{tabular*}
\end{table}

Notably, the averaged velocity values obtained from PIV are by up to 38\% less than those found in the simulations.
These errors can arise from limited tracer density, finite sampling volume, and image resolution \cite{PIV_ERROR}. 
Seeding inhomogeneity and particle clustering can bias local displacement and distort the measured velocity, especially in shear regions and near the droplet interface \cite{antonov2019gas,volkov2018temperature}.

\begin{table}[t!]
\centering
\renewcommand{\arraystretch}{1.5}
\caption{Median absolute error (Eq. \ref{eq:medae}) of radial velocity between the simulations and experimental data after 8 ms. The numerical results are linearly interpolated without any further modifications.}
\label{tab:mae_comparison_original}
\setlength{\tabcolsep}{3pt}
\begin{tabular*}{\columnwidth}{@{\extracolsep{\fill}}ccc@{}}
\hline \hline
\makecell[c]{$We_{imp}$} &
\makecell[l]{Hoffman function \cite{kistler1993hydrodynamics}} &
\makecell[l]{Generalized HVT Law \cite{hoffman1975study}} \\
\hline
20 & 0.026760 & 0.100008 \\
80 & 0.095143 & 0.162777 \\
250 & 0.054199 & 0.266119 \\
\hline \hline
\end{tabular*}
\end{table}

When measuring velocity fields inside a droplet using PIV, significant errors arise in the vicinity of the free surface and, in particular, near the droplet rim \cite{kumar2017internal, castrejon2011dynamics, kang2004quantitative, chaze2017spatially, ribeiro2023insights}. 
These errors are caused by optical distortions due to refraction of light at the curved gas-liquid interface. 
Similar effects have repeatedly been reported in experiments on droplet impact on a surface and on internal-flow visualization.
To deal with the PIV measurement error, the velocity values preprocessing procedure described in Section \ref{subsec:velocity-preproc} was applied.

\begin{figure*}[!t]
    \centering
    \includegraphics[width=1\textwidth]{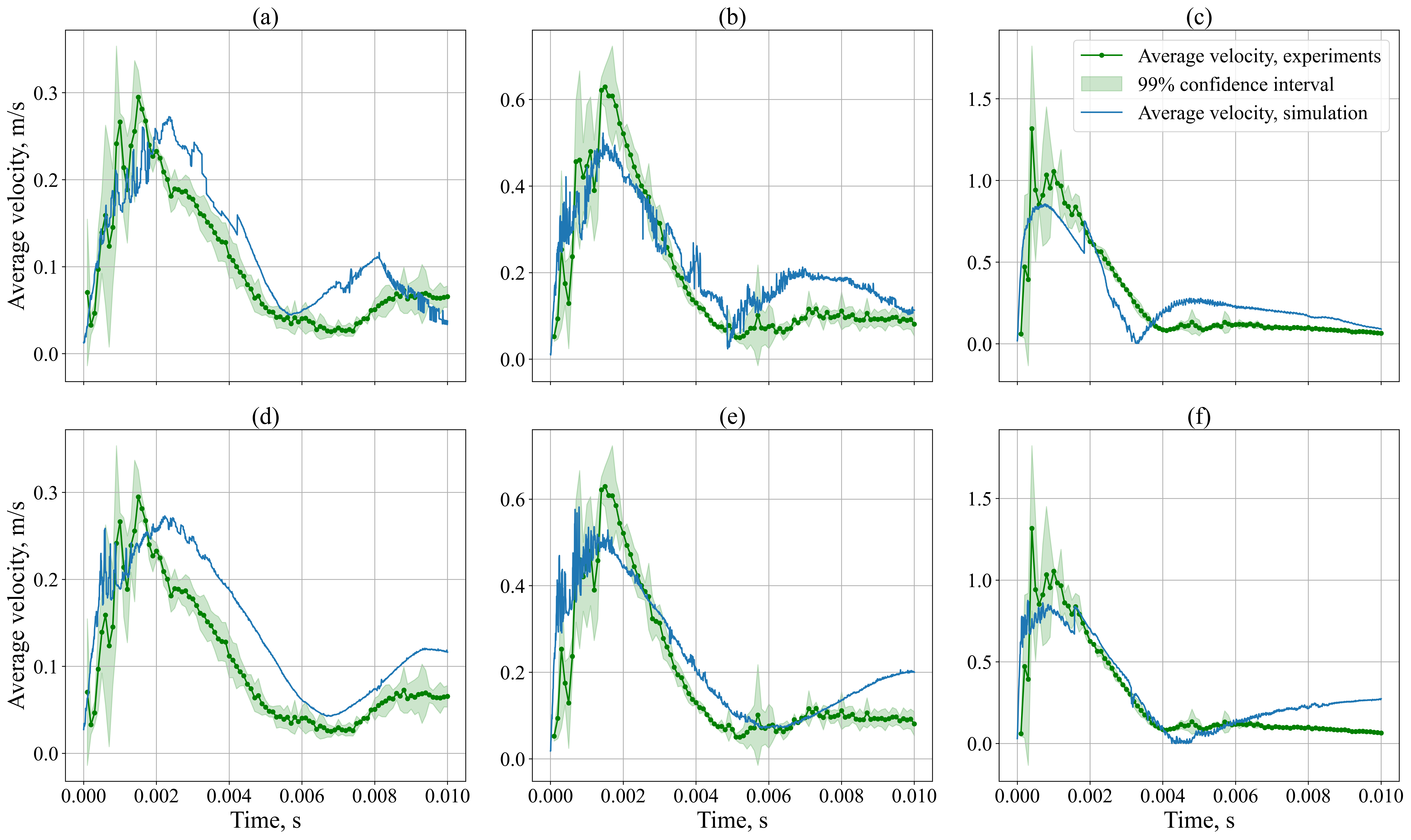}
    \caption{Comparison of the values of the average velocity inside the droplet with experimental data. CFD data are interpolated and filtered as explained in Section \ref{subsec:velocity-preproc}, for consistency with the resolution of PIV. Hoffman function-based simulations are presented at the first row (a, b, c). Simulations with generalized Hoffman-Voinov-Tanner law are presented at the second row (d, e, f). Simulations with $We_{imp}=20$ are presented at the first column of plot (a, d), $We_{imp}=80$ - the second column (b, e), $We_{imp}=250$ - the third column (c, f).}
    \label{fig:velocity_comparison}
\end{figure*}

\subsection{\label{sec:proc-rad-vel}Average radial velocity after PIV correction preprocessing}

\begin{figure*}[t!]
    \centering
    \includegraphics[width=1.0\textwidth]{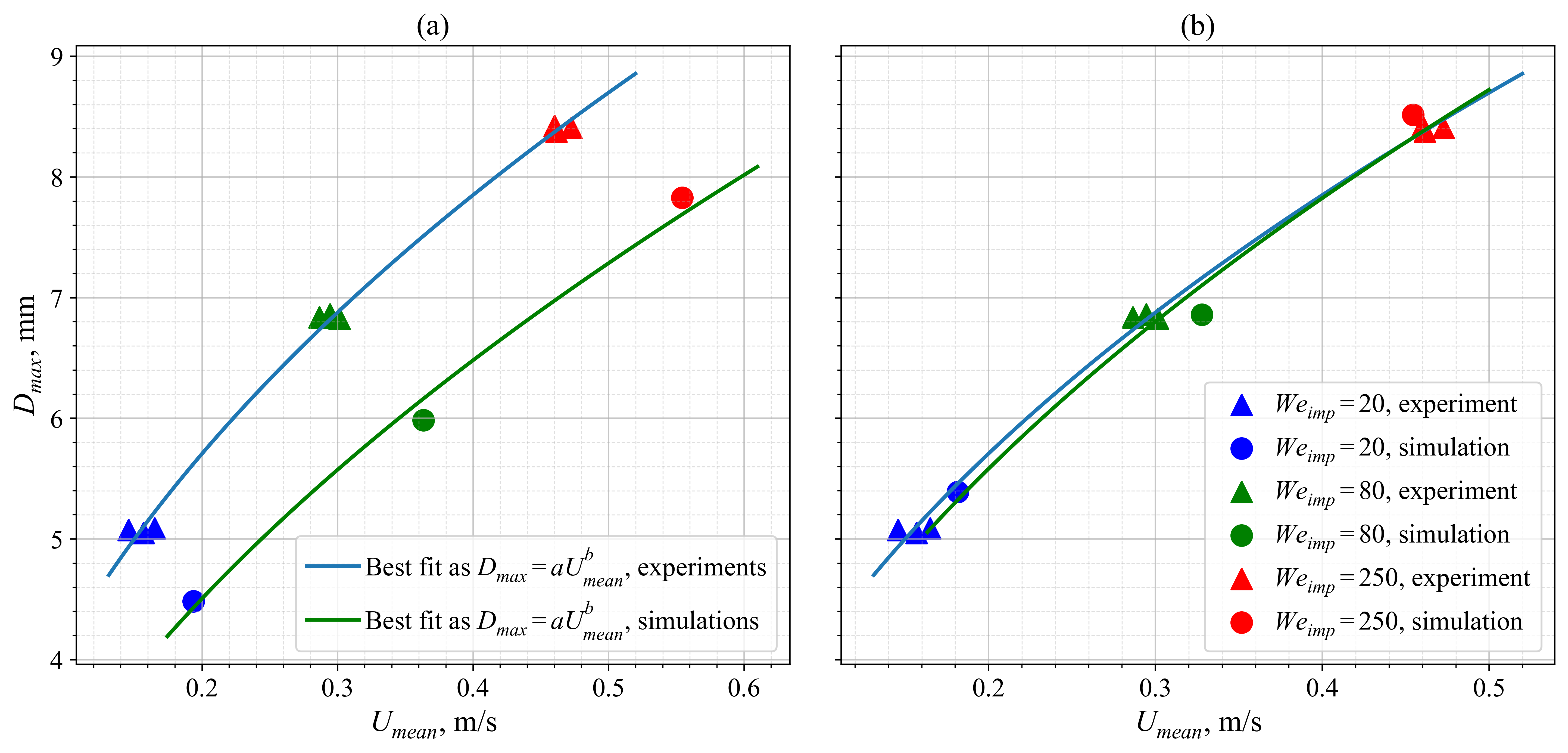}
    \caption{A comparison of numerical and experimental values of the maximum droplet spreading diameter over the mean velocity of internal flows for the period from the droplet-wall contact
and up to the maximum spreading. (a) Hoffman function-based dynamic contact angle model, (b) generalized Hoffman-Voinov-Tanner law. Values of $a$ and $b$ are presented in Table \ref{tab:appr_coeffs}.}
    \label{fig:dmax_umean}
\end{figure*}

\begin{table*}[t!]
\renewcommand{\arraystretch}{1.5}
\centering
\caption{Absolute and relative errors of mean velocity $U_{mean}$ between numerical data and experiments.}
\label{tab:umean_dev}
\begin{tabular}{lp{3.5cm}p{3.5cm}p{2.5cm}p{2.5cm}} 
\hline \hline
$We_{imp}$ & 
\makecell[l]{Absolute error, m/s,\\ Hoffman function \cite{kistler1993hydrodynamics}} & 
\makecell[l]{Absolute error, m/s,\\ generalized HVT law \cite{hoffman1975study}} & 
\makecell[l]{Relative error, \%,\\ Hoffman function} & 
\makecell[l]{Relative error, \%,\\ generalized HVT law} \\ \hline
20  & 0.02872 & 0.01658  & 17.41627 & 10.05298 \\
80  & 0.06896 & 0.03345  & 23.42390 & 11.36375 \\
250 & 0.08186 & -0.00564 & 17.32535 & -1.22512 \\ \hline \hline
\end{tabular}
\end{table*}

To further verify the results of the numerical simulation, the calculated velocity fields were compared with the experimental data obtained by the PIV method. 
To minimize the systematic error of PIV, which is associated with underestimation of measured velocity values, the data were processed with the procedure described in Section \ref{subsec:velocity-preproc}.
The comparison of simulation average velocity values $v_{avg}$ with experimental values is presented in Fig.~\ref{fig:velocity_comparison}.

A comparative analysis of velocity profiles shows that the model based on the generalized Hoffman-Voinov-Tanner law (Fig. \ref{fig:velocity_comparison} (d, e, f)) in the receding phase results in an unrealistic dynamical behavior. 
At the final stage, a monotonous increase in velocity appears, which contradicts dynamics observed in the experiment. 
On the contrary, the model that uses the Hoffman function (Fig. \ref{fig:velocity_comparison} (a, b, c)) better replicates the dynamics of the receding phase. 
The median absolute error (MedAE, Eq. \ref{eq:medae}) between the simulations and the experimental data for the receding phase (the time period after 8 ms from the droplet-wall contact) is listed in Table \ref{tab:mae_comparison}.
The Hoffman function-based simulation demonstrates a significantly better agreement with the experiment across all impact velocities, exhibiting median absolute errors up to three times lower than the generalized Hoffman-Voinov-Tanner Law, particularly at the highest velocity.

\begin{table}[h]
\centering
\renewcommand{\arraystretch}{1.5}
\caption{Median absolute error (Eq. \ref{eq:medae}) of radial velocity between the simulations and experimental data after 8 ms}
\label{tab:mae_comparison}
\setlength{\tabcolsep}{4pt}
\begin{tabular*}{\columnwidth}{@{\extracolsep{\fill}}ccc@{}}
\hline \hline
\makecell[c]{$We_{imp}$} & 
\makecell[l]{Hoffman function \cite{kistler1993hydrodynamics}} & 
\makecell[l]{Generalized HVT Law \cite{hoffman1975study}} \\
\hline
20  & 0.013696 & 0.045810 \\
80  & 0.054071 & 0.088659 \\
250  & 0.024150 & 0.170944 \\
\hline \hline
\end{tabular*}
\end{table}

To conclude, analysis of the radial velocity over time (Fig. \ref{fig:velocity_comparison}) shows that the model based on the Hoffman function reproduces the droplet receding dynamics significantly more accurately, especially at higher velocities.
At the same time, the generalized Hoffman-Voinov-Tanner law leads to a slightly better agreement with experimental data for the maximum droplet spreading diameter and mean velocity during the spreading phase. 
These findings underscore the necessity of a combined validation approach, integrating both geometric parameters (e.g., spreading diameter) and dynamic characteristics (e.g., internal flow velocities), to ensure a robust and physically consistent model of droplet impact dynamics.

\subsection{\label{sec:u_mean}Internal flow velocity effect on maximum spreading diameter}

For a comprehensive validation of the simulation results, the dependence of the maximum spreading diameter $D_{max}$ over the average velocity of the internal flows $U_{mean}$ was observed.
$U_{mean}$ was calculated for the period from the droplet-wall contact and up to the maximum spreading, the same as in the experiment \cite{ASHIKH_COAL}.
The numerical results derived by using the Hoffman function-based model are presented in Fig. \ref{fig:dmax_umean} (a), the simulation data obtained by employing the generalized Hoffman-Voinov-Tanner law is shown in Fig. \ref{fig:dmax_umean} (b).
The numerical points can be described by a power law of the form $D_{max}=a U_{mean}^b$.
The coefficients of the approximation functions (constant $a$ and power $b$) are presented in Table \ref{tab:appr_coeffs}.
All models indicate that the exponent $b$ is close to 0.5, which suggests quadratic scaling of the internal flow $U_{mean} \propto D_{max}^2$.

\begin{table}[t!]
\centering
\renewcommand{\arraystretch}{1.5}
\caption{Approximation coefficients $D_{max} = a U_{mean}^{b}$}
\label{tab:appr_coeffs}
\setlength{\tabcolsep}{4pt}
\begin{tabular*}{\columnwidth}{@{\extracolsep{\fill}}lcc@{}}
\hline \hline
Data & $a$ & $b$ \\
\hline
Experimental data & 11.9659 & 0.4599 \\
Hoffman function \cite{kistler1993hydrodynamics} simulation & 10.4785 & 0.5242 \\
Generalized HVT Law \cite{hoffman1975study} simulation & 12.2308 & 0.4874 \\
\hline \hline
\end{tabular*}
\end{table}

A detailed analysis of the maximum droplet diameter spreading $D_{max}$ is presented in Section \ref{sec:max_drop_spr_diam}.
Most importantly, numerical results obtained using the generalized Hoffman-Voinov-Tanner law show a better accuracy according to the maximum droplet spreading diameter criterion.

Percent and absolute difference of averaged experimental data and numerical results for $U_{mean}$ is presented in Table \ref{tab:umean_dev}.
The comparison of the numerical mean velocity $U_{mean}$ with the experimental values shows that both models systematically overestimate the velocity across $We_{imp}=20$ and $We_{imp}=80$.
For $We_{imp}=20$, the Hoffman function-based values are 17.42\% higher than the experimental data, while the generalized HVT law gives 10.05\% deviation, which is slightly less.
At $We_{imp}=80$, the deviations remain noticeable: 23.42\% for the Hoffman function and 11.36\% for the generalized HVT law.
However, at $We_{imp}=250$, the generalized HVT law demonstrates significantly improved agreement with the experimental value (-1.23\% difference), whereas the Hoffman function still exhibits a notable overprediction (17.33\% difference).
It is important to mention that the best agreement in terms of the mean velocity for generalized Hoffman-Voinov-Tanner law is obtained during the spreading stage.
It was shown earlier in Section \ref{sec:proc-rad-vel} that the receding phase of this model shows less correct dynamics. 

\subsection{\label{sec:vel-field}Velocity fields}

\begin{figure*}[t!]
    \centering
    \includegraphics[width=0.9\textwidth]{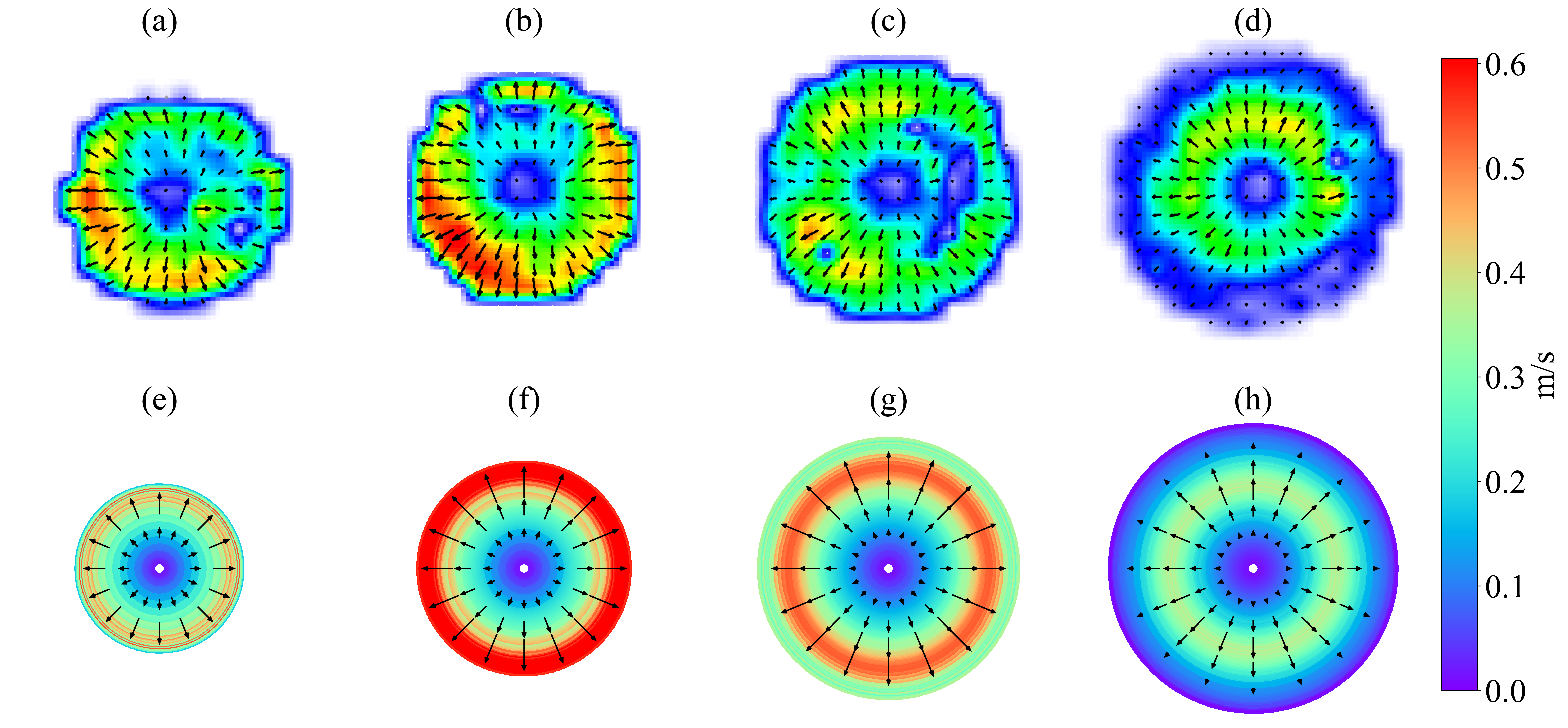}
    \caption{The evolution of velocity fields in the longitudinal section of the spreading droplet at the following time points: (a) 0.001 s; (b) 0.0015 s; (c) 0.0024 s; (d) 0.004 s. The upper row shows the experimental data; the lower one reveals the numerical results obtained with Hoffman function used for simulation of dynamic contact angle.}
    \label{fig:velocity_field}
\end{figure*}

\begin{figure*}[t!]
    \centering
    \includegraphics[width=0.9\textwidth]{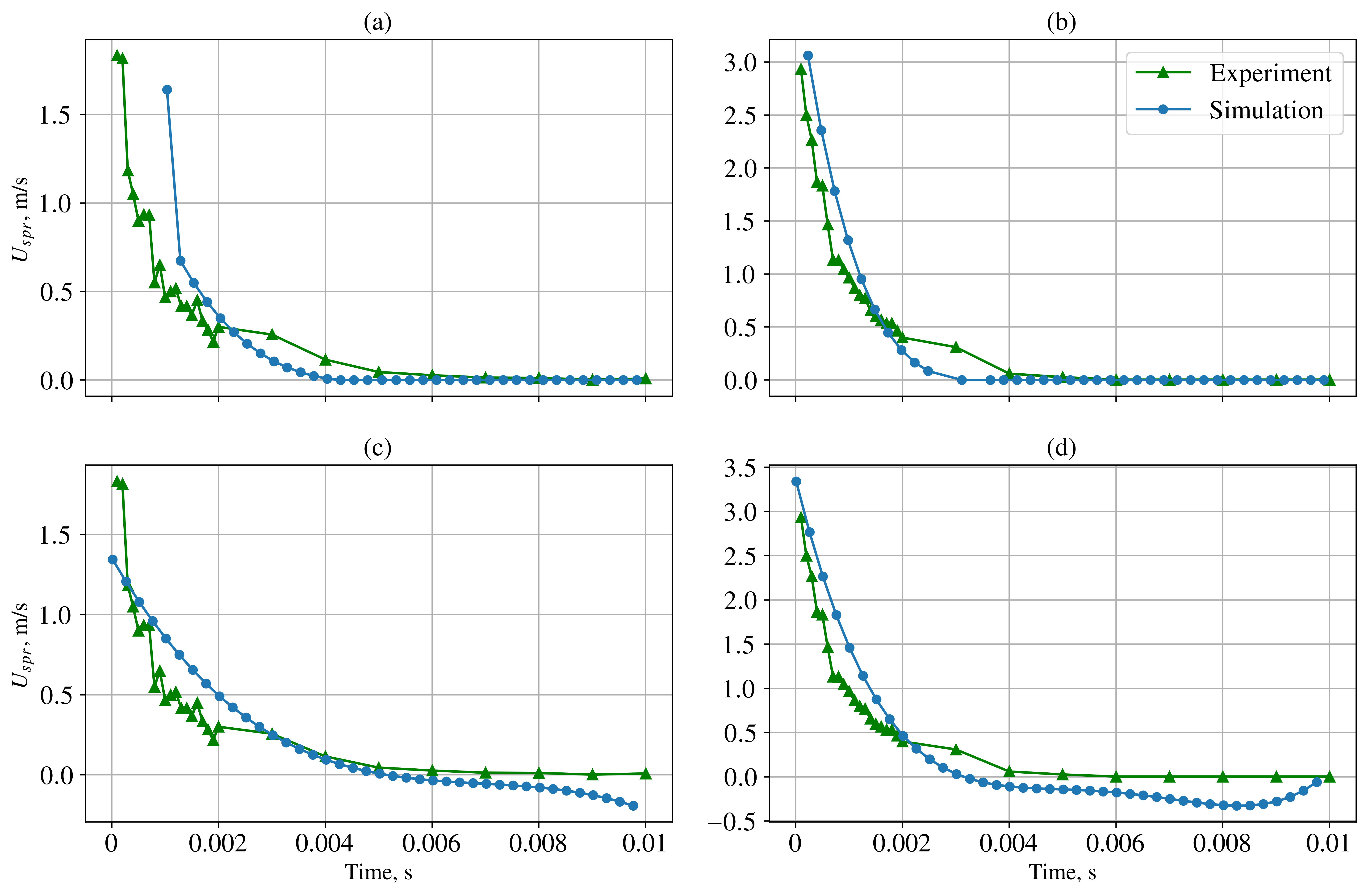}
    \caption{Droplet spreading velocity. First row (a, b) - simulation with Hoffman function, second row (c, d) - simulation with generalized Hoffman-Voinov-Tanner law. First column (a, c) - impact $We_{imp}=20$, second column (b, d) - impact $We_{imp}=250$.}
    \label{fig:u_spr}
\end{figure*}

The evolution of velocity fields in the longitudinal section of a spreading droplet at certain time points is presented in Fig. \ref{fig:velocity_field}.
Experimental data are compared with the numerical results obtained using the Hoffman function-based model at $We_{imp}=20$.
To visualize the velocity fields, the absolute values of the radial velocity were linearly interpolated and rotated around the symmetry axis (the white dot in the center of each subplot). 
The numerical results reproduce the experimental data with good accuracy. 
The main tendencies of velocity values inside the droplet body are reproduced.
However, the calculated velocities closer to the rim of the droplet are higher than those in experiment.
These velocity deviations may be caused by optical distortions near the interface, as described in Section \ref{subsec:velocity-original}.
This effect has been taken into account during average velocity preprocessing (Section \ref{subsec:velocity-preproc}).

\begin{table*}[t!]
    \centering
    \renewcommand{\arraystretch}{1.5}
    \caption{Comparison of contact angle models and their agreement with geometric and kinematic validation parameters.}
    \label{tab:contact_angle_models_comparison}
    \setlength{\tabcolsep}{4pt} 
    \begin{tabular*}{\textwidth}{@{\extracolsep{\fill}}>{\raggedright\arraybackslash}p{2.8cm}>{\raggedright\arraybackslash}p{1.1cm}>{\raggedright\arraybackslash}p{2.4cm}>{\raggedright\arraybackslash}p{3.1cm}>{\raggedright\arraybackslash}p{3.1cm}>{\raggedright\arraybackslash}p{2.3cm}}
        \hline \hline
        Parameter & Stage & Visualisation of the described data &
        Generalized HVT law \cite{hoffman1975study} &
        Hoffman function \cite{kistler1993hydrodynamics} &
        Best model \\
        \midrule
        Percentage difference of the maximum droplet spreading diameter from the experiment &
        Spreading &
        Section \ref{sec:max_drop_spr_diam}: Table \ref{tab:contact_angle_models} &
        $\leq 6.3\%$ &
        $6.84$-$12.52\%$ &
        \textbf{Generalized HVT law} \\
        \hline
        The average radial velocity in the longitudinal section of the droplet, $v_{avg}$ &
        Spreading &
        Section \ref{sec:proc-rad-vel}: Figure \ref{fig:velocity_comparison} &
        Acceptable agreement with experimental data &
        Acceptable agreement with experimental data &
        No clear winner \\
        \cmidrule(lr){2-6}
         &
        Receding &
        Section \ref{sec:proc-rad-vel}: Figure \ref{fig:velocity_comparison}, Table \ref{tab:mae_comparison} &
        The MedAE (Eq.~\ref{eq:medae}) is up to three times higher than for the Hoffman-function-based model &
        Lowest MedAE for all $U_0$ values &
        \textbf{Hoffman function} \\
        \hline
        The average velocity of the internal flow during the spreading stage, $U_\mathrm{mean}$ &
        Spreading &
        Section \ref{sec:u_mean}: Figure \ref{fig:dmax_umean}, Table \ref{tab:umean_dev} &
        Best agreement, with the minimum error in $U_\mathrm{mean}$ &
        Systematic overestimation of $U_\mathrm{mean}$ compared to the generalized HVT law &
        \textbf{Generalized HVT law} \\
        \hline
        Droplet spreading velocity, $U_{\mathrm{spr}}$ &
        Spreading &
        Section \ref{sec:droplet-spr-vel}: Figure \ref{fig:u_spr} &
        There is slightly better agreement with the initial decay of $U_{\mathrm{spr}}$ for low $We_{\mathrm{imp}}$, but greater discrepancies in the peak velocity values &
        Correct overall trend of $U_{\mathrm{spr}}$: a high initial value followed by a monotonic decay towards zero, with a slightly delayed decay at low $We_{\mathrm{imp}}$ &
        No clear winner \\
        \cmidrule(lr){2-6}
         &
        Receding &
        Section \ref{sec:droplet-spr-vel}: Figure \ref{fig:u_spr} &
        Physically less correct behaviour: negative $U_{\mathrm{spr}}$ when the contact line is already pinned &
        Physically consistent dynamics: the contact line is pinned, $U_{\mathrm{spr}} = 0$ &
        \textbf{Hoffman function} \\
        \hline \hline 
    \end{tabular*}
\end{table*}

\subsection{\label{sec:droplet-spr-vel}Droplet spreading velocity}

To explore droplet dynamics, the calculated spreading velocity ($U_{spr}$) was compared to experimental results.
The calculated $U_{spr}$ values for cases $We_{imp}=20$ (Fig. \ref{fig:u_spr} (a, c)) and $We_{imp}=250$ (Fig. \ref{fig:u_spr} (b, d)) with the use of the Hoffman function (Fig. \ref{fig:u_spr} (a, b)) and the generalized Hoffman-Voinov-Tanner law (Fig. \ref{fig:u_spr} (c, d)) show different results compared with experiments.

For both impact conditions, the Hoffman function-based model reproduces the qualitative structure of the spreading dynamics. 
An initially high spreading velocity followed by a rapid monotonic decay toward zero as the lamella slows down is observed.
This is consistent with the expected physical behavior of the real droplet. 
In particular, at later times the Hoffman function-based model correctly predicts that the contact line motion approaches pinning. 
The main limitation of the Hoffman model is that, at an early stage for $We_{imp}=20$ (Fig. \ref{fig:u_spr} (a)), the numerical curve shows a decay in spreading velocity that occurs slightly later than in the experimental data.

The generalized HVT law shows a contrastingly different behavior.
In the early stage ($\leq 2~\text{ms}$) its prediction agrees even better with the experimental data for lower $We_{imp}$ (Fig.~\ref{fig:u_spr}~(c)).
This indicates that the generalized HVT law can accurately describe droplet dynamics immediately after impact.
However, at a later stage, the generalized HVT law develops a less physical trend. 
After the spreading velocity decays toward zero, the model continues past zero and $U_{spr}$ produces negative values (Fig. \ref{fig:u_spr} (c, d)). 
The model predicts that the contact line is receding even though the measured contact line has already stabilized. 
This behavior is not observed experimentally and should not occur in a physically consistent simulation.

Thus, despite its higher early-time accuracy, the generalized HVT law fails to enforce the correct asymptotic constraint $U_{spr}\geq 0$ during the receding phase, and therefore does not reproduce the correct dynamics of the droplet. 
By contrast, the Hoffman function-based approach, while less precise in the very initial stage, preserves physically more correct behavior throughout all stages.

This comparison highlights a key methodological point: to adequately model the droplet-wall impact including the kinematic phase and the phases of spreading and receding, it is necessary to take into account not only the diameter of spreading, but also the velocity distribution inside the droplet. 
Using the maximum spreading diameter as the only criterion for validating numerical results and the correctness of the model is a methodologically weaker approach.
The reason is that, even with minimal discrepancies in the maximum spreading diameter, significant discrepancies can occur during the receding stage, up to a qualitative change in the dynamics of the process. 
In particular, there may be cases when there is no receding at all, which indicates fundamental differences in the behavior of the system. 
Thus, for a full-fledged validation of the model, a comprehensive analysis of both geometric parameters (e.g., the spreading diameter) and dynamic characteristics (e.g., internal flow velocities) is required.

\subsection{Combined dynamic contact angle model}

When comparing the results of two dynamic contact angle models (Section \ref{sec:max_drop_spr_diam} - \ref{sec:droplet-spr-vel}), the generalized HVT law \cite{hoffman1975study} is superior during the spreading stage, as it exhibits a lower deviation in the maximum droplet spreading diameter and the average velocity of the internal flow ($U_{mean}$).
On the other hand, the Hoffman function-based model \cite{kistler1993hydrodynamics} is superior in terms of the accuracy of dynamics during receding stage ($v_{avg}, U_{spr}$).

Our comparison of the two models is presented in Table \ref{tab:contact_angle_models_comparison}.
The spreading and the receding phases are compared separately to highlight the distinct predictive capabilities of each model at different stages of droplet evolution.
The generalized Hoffman-Voinov-Tanner law performs better at the spreading stage, providing the closest agreement with experiment for the maximum droplet spreading diameter and for the mean internal-flow velocity during spreading. 
However, it remains significantly less accurate during the receding stage than the Hoffman function-based model, even despite its smaller deviation in the maximum spreading diameter. 
In contrast, the Hoffman function-based model systematically outperforms the generalized Hoffman-Voinov-Tanner law for kinematic metrics characterizing the receding phase, such as the average radial velocity and the droplet spreading velocity, and it preserves a physically consistent contact-line behavior. 
For some parameters (e.g., the average radial velocity during spreading and the early-time evolution of the spreading velocity), no single model clearly dominates.

In order to unite the advantages of both DCA models, we suggest a combined model for the advancing stage

\begin{equation}
     \theta_D^3-\theta_0^3=8.78 Ca, \quad Ca > 0,
    \label{ref:comb_adv}
\end{equation}

and the receding stage

\begin{equation}
     \theta = \frac{\theta_0}{\pi - \theta_0} \left[\theta_0 - f_H\left(-Ca + f_H^{-1}(\theta_0)\right)\right] + \theta_0, \quad Ca < 0.
    \label{ref:comb_rec}
\end{equation}

\begin{figure*}[t!]
    \centering
    \includegraphics[width=1.0\textwidth]{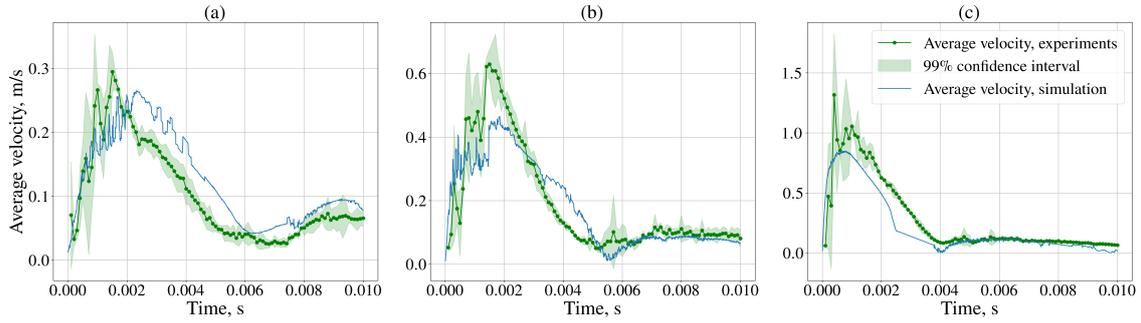}
    \caption{Comparison of the values of the average velocity inside the droplet with experimental data. The plots represent the results of the combined dynamic contact angle model. CFD data are interpolated and filtered as explained in Section \ref{subsec:velocity-preproc}, for consistency with the resolution of PIV. Simulation with $We_{imp}=20$ is presented at the first plot (a), $We_{imp}=80$ - the second plot (b), $We_{imp}=250$ - the third plot (c).}
    \label{fig:velocity_comparison_combined}
\end{figure*}

\begin{figure*}[t!]
    \centering
    \includegraphics[width=0.9\textwidth]{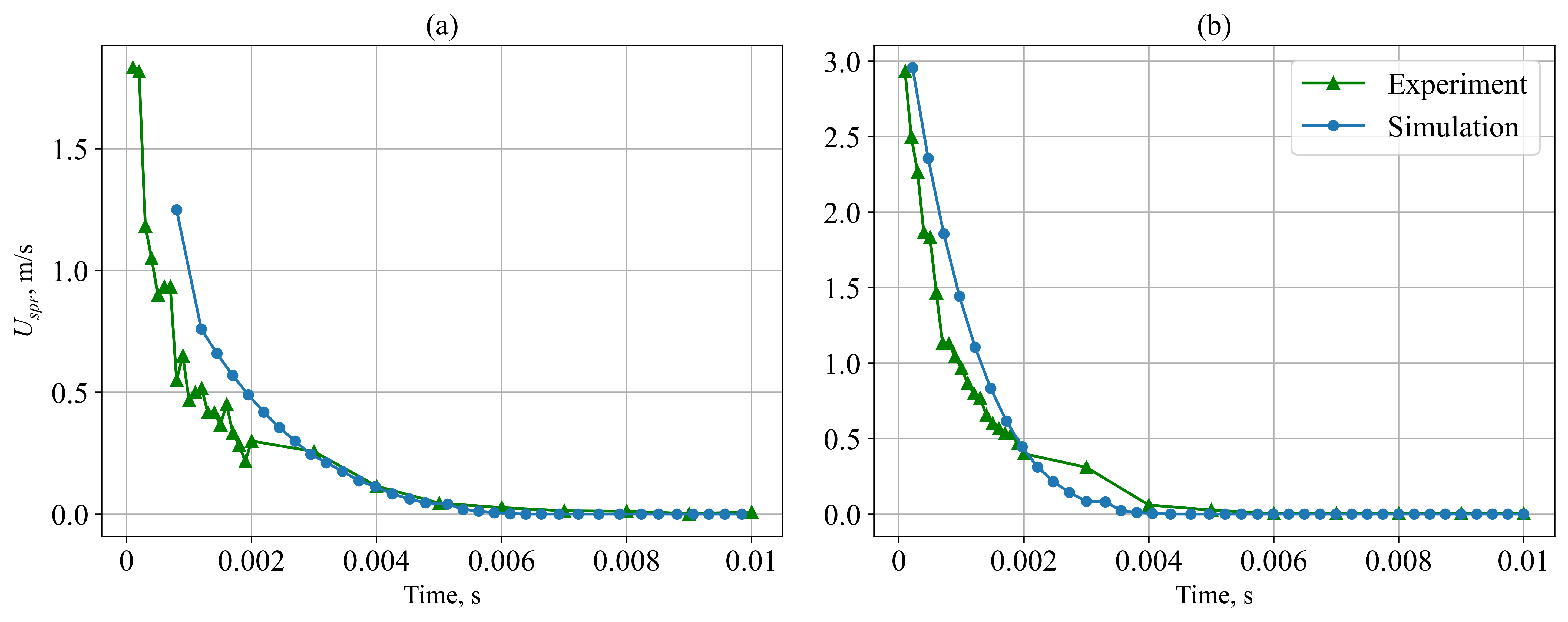}
    \caption{Droplet spreading velocity. Plots represent results for combined dynamic contact angle model. First plot (a) - $We_{imp}=20$, second plot (b) - $We_{imp}=250$.}
    \label{fig:u_spr_combined}
\end{figure*}

During the spreading stage ($Ca>0$), the generalized Hoffman-Voinov-Tanner law is used, employing the previously selected constant $\kappa = 8.78$ (Section \ref{sec:max_drop_spr_diam}).
The Hoffman function-based DCA model is used during receding stage ($Ca<0$).
This combined DCA model was applied with user-defined function.

The maximum droplet spreading diameter deviation is the same as for the generalized Hoffman-Voinov-Tanner law with the selected constant $\kappa = 8.78$ (Table \ref{tab:contact_angle_models}).
The deviation from experiments does not exceed 7\%.
This result outperforms Hoffman function-based model for every $We_{imp}$.

The comparison of average radial velocity values ($v_{avg}$) for combined DCA with experimental data is presented in Fig. \ref{fig:velocity_comparison_combined}.
The values were processed with the procedure described in Section \ref{subsec:velocity-preproc}.
The velocity profiles for the spreading stage are almost identical to those obtained using the generalized Hoffman-Voinov-Tanner law (Fig. \ref{fig:velocity_comparison} (a, b, c)).
However, the receding stage shows more accurate results comparing to generalized HVT law, especially for $U_0=0.63$ m/s and $U_0=1.17$ m/s (Fig. \ref{fig:velocity_comparison_combined} (a, b)).
Table \ref{tab:mae_comparison_combined} reports the median absolute error (MedAE, Eq. \ref{eq:medae}) between simulation and experimental data for the receding phase (after 8 ms from droplet-wall impact).
For all cases combined DCA shows better results compared to Hoffman function-based data.

Droplet spreading velocity for simulations using combined DCA is presented in Fig. \ref{fig:u_spr_combined}.
Across both impact regimes, the model based on the combined DCA successfully reproduces the overall pattern of droplet spreading.
It predicts a brief initial phase of very fast spreading, after which the velocity rapidly and monotonically declines toward zero as the lamella decelerates.
This evolution matches the physically expected behavior of a real droplet.
Crucially, at later stages the combined DCA-based model also correctly captures the tendency of the contact line motion to slow down and approach pinning, in contrast to the generalized Hoffman-Voinov-Tanner law (Fig. \ref{fig:u_spr} (c, d)).

\begin{table}[t!]
\centering
\renewcommand{\arraystretch}{1.5}
\caption{Median absolute error (Eq. \ref{eq:medae}) of radial velocity between the simulations and experimental data after 8 ms. The lowest error for every $We_{imp}$ is highlighted in bold font.}
\label{tab:mae_comparison_combined}
\setlength{\tabcolsep}{3pt}
\begin{tabular*}{\columnwidth}{@{\extracolsep{\fill}}cccc@{}}
\hline \hline
\makecell[c]{$We_{imp}$} &
\makecell[l]{Combined} &
\makecell[l]{Hoffman function \cite{kistler1993hydrodynamics}} &
\makecell[l]{Generalized \\ HVT Law \cite{hoffman1975study}} \\
\hline
20 & \textbf{0.021770} & 0.026760 & 0.100008 \\
80 & \textbf{0.014934} & 0.095143 & 0.162777 \\
250 & \textbf{0.032246} & 0.054199 & 0.266119 \\
\hline \hline
\end{tabular*}
\end{table}

\begin{figure}[t!]
    \centering
    \includegraphics[width=0.45\textwidth]{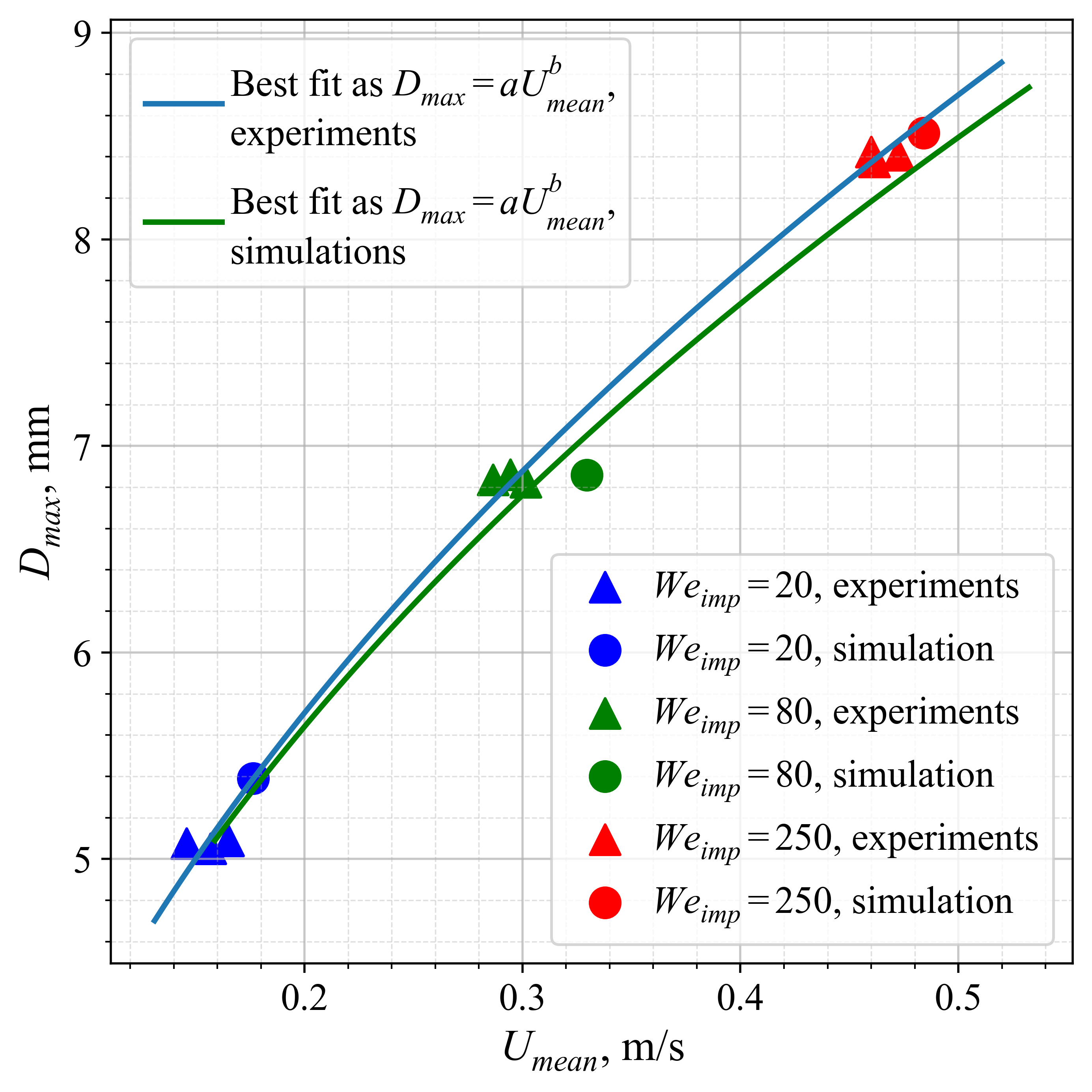}
    \caption{Comparison of the values of the average velocity inside the droplet with experimental data. Plots represent results for combined dynamic contact angle model. The CFD data fitting coefficient values are $a=11.5756$ and $b=0.4466$.}
    \label{fig:dmax_umean_combined}
\end{figure}

The effect of the internal flow velocities on maximum droplet spreading diameter is presented in Fig. \ref{fig:dmax_umean_combined}.
The points of the simulation results are close enough to the experimental data.
The observed numerical results may be presented in form of a power law. 

The comparison of relative error for all DCA models of $U_{mean}$ is presented in Table \ref{tab:umean_dev_combined_corrected_final}.
Combined DCA is superior for $We_{imp}=20$ and has almost the same results as generalized HVT law for other cases.
For all simulations, combined DCA shows better results than the Hoffman function-based simulation.

To test the generalization capabilities of the proposed combined DCA model, a number of additional simulations were performed.
Simulation setups of these calculations are presented in Table \ref{tab:combined_dca_add_sim}.
The density of the liquid is the same for all simulations, $\rho=1154$~kg/m$^3$.

\begin{figure*}[t!]
    \centering
    \includegraphics[width=1\textwidth]{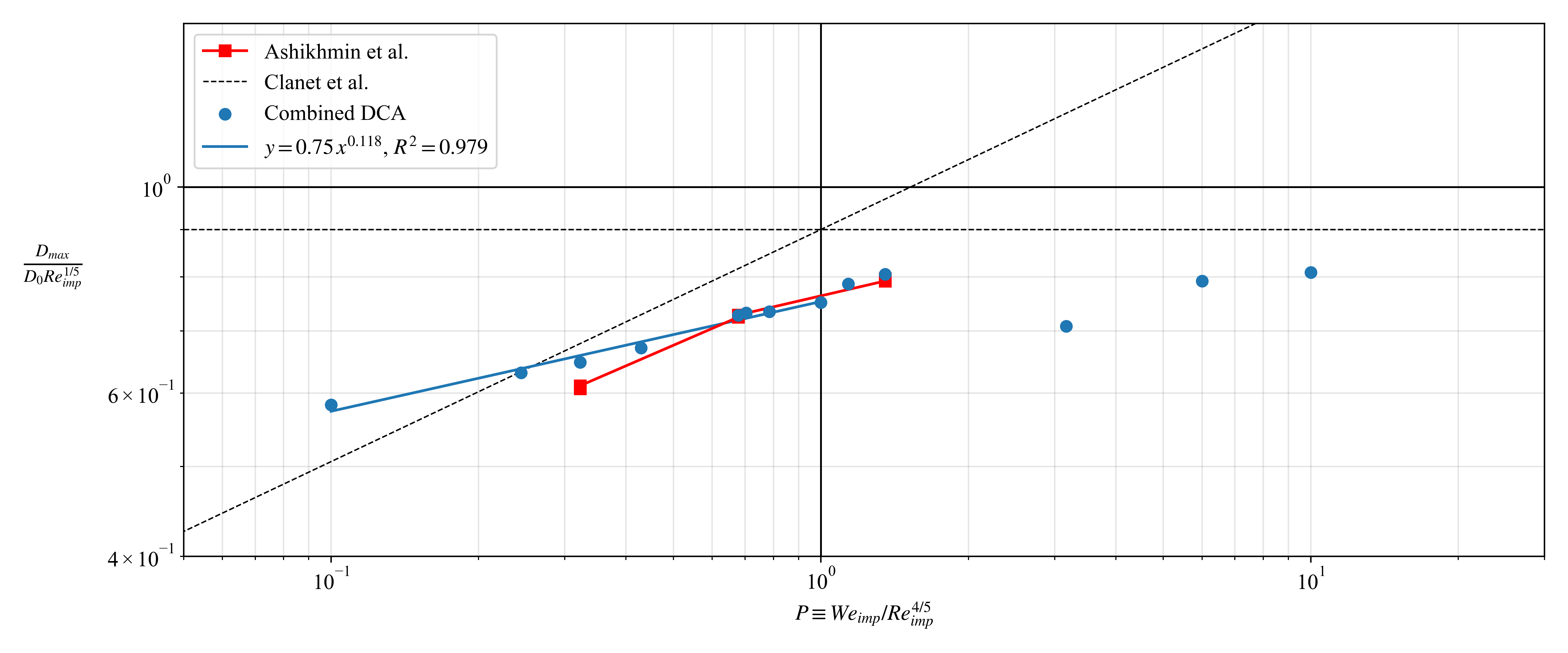}
    \caption{Comparison of numerical results obtained with combined DCA (blue dots) with results obtained by Clanet et al.\cite{clanet2004maximal} (dashed lines) and Ashikhmin et al.\cite{ASHIKH_COAL} (red dots and lines).
    Blue line represents linear trend of numerical results for the capillary regime.}
    \label{fig:clanet_comparison}
\end{figure*}

Numerical results were compared with the dimensionless scaling law reported by Clanet et al. \cite{clanet2004maximal} and with experimental data from Ashikhmin et al. \cite{ASHIKH_COAL}.
The results are presented in Fig. \ref{fig:clanet_comparison}.
The transition between the capillary and viscous regimes occurs near $P \approx 1$, consistent with the original study \cite{clanet2004maximal}.
In the capillary regime, the numerical results follow a power law $P^{0.118}$ with a good fit ($R^2 = 0.979$).
However, the fitted exponent differs from the original scaling law $D_{max}/(D_0 Re_{imp}^{1/5}) \sim P^{1/4}$, where the impact Reynolds number is $Re_{imp}=\rho U_0 D_0 / \mu$.
This may be an artifact of the proposed combined DCA model.

\begin{table}[t!]
\centering
\renewcommand{\arraystretch}{1.5}
\caption{Relative errors of mean velocity $U_{mean}$ between numerical data and experiments. The lowest error for every $We_{imp}$ is highlighted in bold font.}
\label{tab:umean_dev_combined_corrected_final}
\setlength{\tabcolsep}{3pt}
\begin{tabular*}{\columnwidth}{@{\extracolsep{\fill}}cccc@{}}
\hline \hline
\makecell[c]{$We_{imp}$} &
\makecell[c]{Rel. error, \%,\\ combined} &
\makecell[c]{Rel. error, \%,\\ Hoffman function} &
\makecell[c]{Rel. error, \%,\\ generalized HVT\\law} \\
\hline
20  & \textbf{6.978453}  & 17.41627  & 10.05298 \\
80  & 11.953093         & 23.42390  & \textbf{11.36375} \\
250 & 2.433364          & 17.32535  & \textbf{-1.22512} \\
\hline \hline
\end{tabular*}
\end{table}

To obtain a generalized correlation between $D_{max}$ and $U_{mean}$, a dimensionless dependence of the droplet spreading factor $\beta_{max}$ (Eq. \ref{eq:beta}) on the characteristic capillary number $Ca_{char}$ (Eq. \ref{eq:ca_char}) was determined.
\begin{align}
    {\beta_{max}=D_{max}/D_0}, \label{eq:beta} \\
    {Ca_{char}=\mu U_{mean} / \sigma}. \label{eq:ca_char}
\end{align}

\begin{table}[t!]
\renewcommand{\arraystretch}{1.5}
\centering
\caption{Parameters of the additional simulations using the combined DCA model.}
\label{tab:combined_dca_add_sim}
\begin{tabular*}{\columnwidth}{@{\extracolsep{\fill}} c c c c}
\hline \hline
$U_0$, m/s & $D_0$, mm & $\mu$, Pa $x$ s & $\sigma$, N / m \\
\hline
0.50 & 2.9 & 0.010800 & 0.06058 \\
0.80 & 2.9 & 0.010800 & 0.06058 \\
1.80 & 2.9 & 0.010800 & 0.06058 \\
1.17 & 3.5 & 0.010800 & 0.06058 \\
1.17 & 2.9 & 0.012960 & 0.06058 \\
1.17 & 2.9 & 0.000987 & 0.06058 \\
1.17 & 2.9 & 0.017558 & 0.06058 \\
1.17 & 2.9 & 0.164900 & 0.06058 \\
1.17 & 2.9 & 0.312240 & 0.06058 \\
1.17 & 2.9 & 0.010800 & 0.01296 \\
\hline \hline
\end{tabular*}
\end{table}

$Ca_{char}$ is the characteristic capillary number, based on the mean internal flow velocity in the horizontal plane at a height of 300 µm above the substrate ($U_{mean}$), calculated from the moment of impact until the maximum diameter is reached. 
It quantitatively defines the ratio of viscous dissipation to capillary forces.
The dimensionless parameter $Ca_{char}$ indicates how strongly viscosity can affect the flow inside a droplet compared to the influence of surface tension.
As the characteristic velocity in $Ca_{char}$, we use $U_{mean}$, since it enables an assessment of dissipation within a specified region, whereas $U_0$ does not represent the characteristic flow velocities inside the droplet after impact with the substrate.

\begin{figure*}[t!]
    \centering
    \includegraphics[width=1\textwidth]{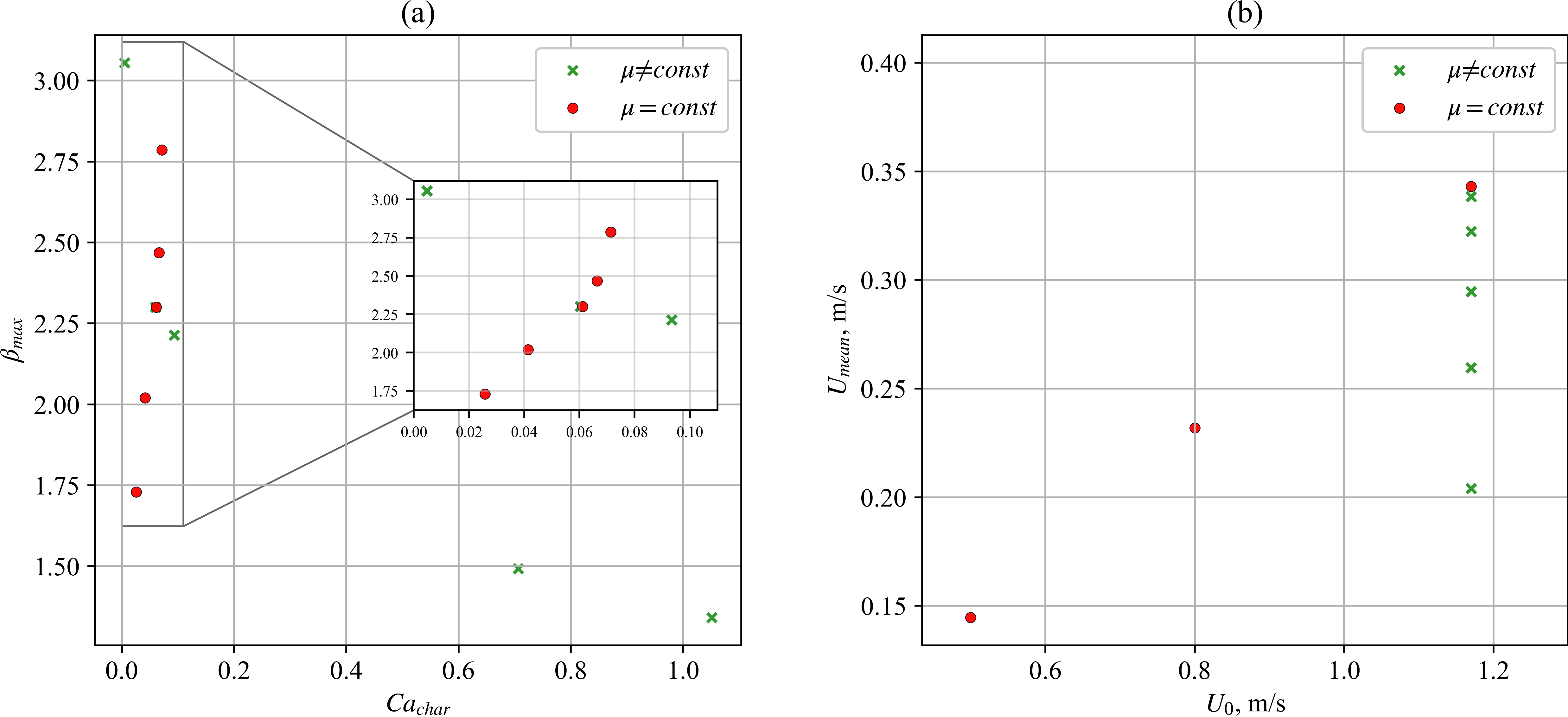}
    \caption{(a) - Droplet spreading ratio $\beta_{max}$ versus characteristic capillary number $Ca_{char}$ with an inset showing low $Ca_{char}$. (b) - Mean internal flow velocity $U_{mean}$ versus droplet impact velocity $U_0$. Data are divided into two series: $\mu=const$ (green dots) and $\mu \neq const$ (red dots).}
    \label{fig:bmax_vs_ca}
\end{figure*}

A comparison between the maximum spreading factor $\beta_{max}$ and the number $Ca_{char}$ for the results obtained using the combined DCA (Table \ref{tab:combined_dca_add_sim}) is shown in Figure \ref{fig:bmax_vs_ca}.
For the analysis of the observed trends, the data are divided into two series.

In the first series of experiments, the liquid viscosity is constant ($\mu=const$), while the impact velocity $U_0$ is the parameter being varied.
In this case, the values of $Ca_{char}$ remain small (see the inset in Fig.~13(a)).
Here, an increase in $\beta_{max}$ is observed simultaneously with an increase in $Ca_{char}$.
This is because increasing $U_0$ leads to an increase in the mean velocity of internal flows $U_{mean}$ (Fig.~13(b)).
In this regime, the spreading process is governed predominantly by inertial forces, $We_{imp}\in[13,\,354]$.

In the second series, where at fixed velocity $U_0$ the viscosity is varied ($\mu \neq const$), a different trend is observed: $b_{\max}$ decreases as $Ca$ increases (Fig. \ref{fig:bmax_vs_ca} (a)).
Physically, this is explained by the fact that as viscosity increases, the contribution of viscous dissipation grows, which suppresses spreading and reduces $\beta_{max}$.

The data presented in Fig. \ref{fig:bmax_vs_ca} make it possible to consider the feasibility of combined dynamic-kinematic validation based on the parameters $\beta_{max}$ and $Ca_{char}$.
The observed relationship between these parameters suggests that, given a sufficient amount of data, the geometric characteristics of the contact line can be used to estimate the internal kinematic properties of the flow.
However, further development of this approach requires expanding the experimental database and obtaining data beyond the ranges considered in the work of Ashikhmin et al \cite{ASHIKH_COAL}.

\newpage
Thus, combined DCA unites the best results of validation of dynamic-kinematic parameters of both models:
    \begin{enumerate}
        \item The deviation of the maximum droplet spreading diameter from the experimental results is minimal.
        \item The correspondence of the radial velocity ($v_{avg}$) to the experiments during the spreading phase is comparable to the Hoffman function-based model (the most accurate model).
        \item The correspondence of the radial velocity ($v_{avg}$) to the experiments during the receding phase is superior.
        \item The relative error for $U_{mean}$ is comparable to the generalized HVT law (the most accurate model) in most cases.
        \item The droplet spreading velocity $U_{spr}$ repeats the trendline of the experimental data and is similar to Hoffman function-based model (the most accurate model, Fig. \ref{fig:u_spr} (a, b)).
    \end{enumerate}

\section{Conclusion}

A water-glycerol droplet impact on sapphire glass was simulated in Ansys Fluent and validated against experiments~\cite{ASHIKH_COAL} for $We_{imp}=20$-$250$.
Two dynamic contact angle models implemented via UDF were compared: the generalized Hoffman-Voinov-Tanner law and the Hoffman function.
A Python preprocessing algorithm for extracting kinematic characteristics and correcting PIV-related measurement errors was introduced.

Validation based solely on the maximum spreading diameter was shown to be insufficient.
Good agreement in $\beta_{\max}$ can accommodate with incorrect receding dynamics.
The generalized Hoffman-Voinov-Tanner law can reproduce $\beta_{\max}$ with small deviations ($\leq 7\%$).
However, it misrepresents the receding stage, producing nonphysical internal-flow acceleration and unrealistic contact-line behavior.
In contrast, the Hoffman-function-based model provides the same geometric accuracy while more consistently reproducing the time evolution of internal velocities and the physically correct trend of spreading rate.
These findings motivate an integrated validation procedure combining geometric and dynamic indicators.
Recommended metrics include $\beta_{\max}$, average internal radial velocity, velocity-field evolution, and droplet spreading velocity.
Overall, this combined validation strategy improves the physical fidelity of the simulations and provides a basis for more informed model selection for spraying, cooling, and related applications.
Based on the complementary strengths of the two models, a combined DCA is proposed.
It uses the generalized Hoffman-Voinov-Tanner formulation during spreading and the Hoffman function-based formulation during receding.
It preserves accurate prediction of $\beta_{\max}$ while ensuring physically consistent receding stage dynamics.

To examine the spreading final states in conjunction with the internal dynamics, we introduce a diagram that relates two dimensionless parameters: $\beta_{max}$ and $Ca_{char}$.
The latter parameter is defined using the mean internal flow velocity in the horizontal plane at a height of 300 µm above the substrate.
$Ca_{char}$ thus defined quantifies the ratio of viscous dissipation to capillary forces.
We conjecture that, given sufficient data, the geometric characteristics of the contact line can be used to estimate the internal kinematic properties of the flow.
Testing this conjecture is a prospective direction for future work on this topic.


\section*{Acknowledgment}

The study was supported by a grant from the Russian Science Foundation No. 23-71-10081 (https://rscf.ru/project/23-71-10081/).

\section*{Data Availability Statement}

The data that support the findings of this study are available from the corresponding author upon reasonable request.



	
\end{document}